\documentclass[nohyper]{JHEP3}

\usepackage{epsfig}
\usepackage{amsfonts}
\usepackage{amsmath}
\usepackage{amssymb}
\usepackage{amsbsy}
\usepackage{latexsym}
\usepackage{epsf}
\usepackage{bbm}
\usepackage{graphicx}
\usepackage{cite}
\usepackage{slashed}

\title{Q-instantons}
\author{E.A. Bergshoeff$^1$, J. Hartong$^1$, A. Ploegh$^1$ and D. Sorokin$^2$
\\
$^1$ Centre for Theoretical Physics, University of Groningen,\\
Nijenborgh 4, 9747 AG Groningen, The Netherlands \\
{\upshape\ttfamily e.a.bergshoeff, j.hartong, a.r.ploegh@rug.nl}
\\ \\
$^2$ INFN, Sezione di Padova\\
via F. Marzolo 8, 35131 Padova, Italia \\
 {\upshape\ttfamily sorokin@pd.infn.it}\\}

\abstract{We construct the half-supersymmetric instanton solutions
that are electric-magnetically dual to the recently discussed
half-supersymmetric Q7-branes. We call these instantons
``Q-instantons''. Whereas the D-instanton is most conveniently
described using the RR axion $\chi$ and the dilaton $\phi$, the
Q-instanton is most conveniently described using a different set of
fields ($\chi', T)$, where $\chi'$ is an axionic scalar. The real
part of the Q-instanton on-shell action is a function of $T$ and the
imaginary part is linear in $\chi'$. Discrete shifts of the axion
$\chi'$ correspond to $PSL(2,\mathbb{Z})$ transformations that are
of finite order. These are \emph{e.g.} pure S-duality
transformations relating weak and strongly coupled regimes. We argue
that near each orbifold point of the quantum axion-dilaton moduli
space $\{\tau\mid\tau\in{{PSL(2,\mathbb{R})}\over{SO(2)\times
PSL(2,\mathbb{Z})}}\}$ the higher order $\mathcal{R}^4$ terms in the
string effective action contain contributions from an infinite sum
of single multiply-charged instantons with the Q-instantons
corresponding to the orbifold points $\tau=i,\rho$.}

\preprint{
\small
UG-08-02\\
 }

\begin{document}

\section{Introduction}

Recently, 7-brane configurations have been investigated with an
emphasis on their supersymmetry properties \cite{Bergshoeff:2006jj}
and their coupling to the bulk IIB supergravity fields
\cite{Bergshoeff:2007aa}. As shown in \cite{Bergshoeff:2006jj}
generic 7-brane configurations contain 7-branes that are associated
to various $SL(2,\mathbb{Z})$ conjugacy classes. The
$SL(2,\mathbb{Z})$ conjugacy classes have been classified in
\cite{DeWolfe:1998eu} and \cite{DeWolfe:1998pr}. In the classical
theory, \emph{i.e.} if one does not take into account charge
quantization there are three families of $SL(2,\mathbb{R})$
conjugacy classes depending on whether $\text{det}\,Q<0$,
$\text{det}\,Q=0$ or $\text{det}\,Q>0$ where $Q$ is such that
$e^Q\in SL(2,\mathbb{R})$. The $(p,q)$ 7-branes correspond to the
case $\text{det}\,Q=0$ whereas the Q7-branes have $\text{det}\,Q>0$
(no 7-branes correspond to $\text{det}\,Q<0$). For each of these
7-branes one can define an axion, that we denote by $\chi'$, with
respect to which the 7-brane is magnetically charged. When we
consider charge quantization the $SL(2,\mathbb{Z})$ conjugacy
classes with $\text{det}\,Q=0$ are given by $\pm T^n$ (with
$n=0,1,2,\ldots$), while for $\text{det}\,Q>0$ they are given by
$S$, $-S$, $(T^{-1}S)^{\pm 1}$ and $(-T^{-1}S)^{\pm 1}$, where $T$
and $S$ transform the axion-dilaton $\tau=\chi+ie^{-\phi}$ as
$T\tau=\tau+1$ and $S\tau=-1/\tau$, respectively. In the notation of
\cite{Bergshoeff:2006jj,Bergshoeff:2007aa} the $SL(2,\mathbb{Z})$
conjugacy classes $-S$ and $-T^{-1}S$ correspond to a single
positive tension Q7-brane. In table \ref{conjugacyclassesandbranes}
we give an interpretation of the $\text{det}\,Q=0$ and
$\text{det}\,Q>0$ $SL(2,\mathbb{Z})$ conjugacy classes in terms of
D7-branes as well as in terms of positive tension Q7-branes.

\begin{table}
\begin{center}
\begin{tabular}{|c|c|}
  \hline
  $SL(2,\mathbb{Z})$ conj. class. & branes \\
  \hline
  $T^n$ & $n$ D7-branes \\
  &\\
  $(-S)^n\,,$ $n\le 4$ & $n$ $(-S)$-branes\\
  &\\
  $(-T^{-1}S)^n\,,$ $n\le 6$ & $n$ $(-T^{-1}S)$-branes\\
  \hline
\end{tabular}
\end{center}
\caption{$SL(2,\mathbb{Z})$ conjugacy classes and branes;
$n\in\mathbb{N}$.}\label{conjugacyclassesandbranes}
\end{table}

The objects that are electrically charged under $\chi'$ and that are
dual to a positive tension Q7-brane are dubbed ``Q-instantons''.
These are half-BPS solutions of Euclidean IIB supergravity. It is
well-known that the object that is dual to the D7-brane is the
so-called D-instanton \cite{Gibbons:1995vg}. In this paper we
present a path integral analysis of the Q-instantons providing us
with their tunneling interpretation and we derive their basic
physical properties such as their charge and on-shell Euclidean
action.

At first sight the existence of new half-supersymmetric instanton
solutions to Euclidean IIB supergravity might be surprising since a
simple analysis of the Killing spinor equations and field equations
seems to lead to the unique D-instanton solution of
\cite{Gibbons:1995vg}, up to an $SL(2,\mathbb{Z})$ transformation of
the D-instanton into a $(p,q)$-instanton. There remains however the
possibility that there exist instantonic solutions that differ from
the D-instanton due to a difference in the source and boundary term.
This is in fact what happens and it leads to different on-shell
actions for the D- and Q-instantons.

In \cite{Green:1997tv} it was shown that the D-instanton contributes
to higher order corrections to the string effective action of the
form of $\mathcal{R}^4$ terms. Since Q-instantons preserve the same
supersymmetries as the D-instanton, they are expected to contribute
to the same $\mathcal{R}^4$ terms as well. We will argue that this
is the case.

The paper is organized as follows. In Section
\ref{sec:IIBconjugacyclasses} we first discuss the general idea of a
Q-brane in type IIB supergravity. In Section \ref{sec:Dinstanton} we
review the construction and properties of the D-instanton. These
results are compared in Section \ref{sec:Euclideanaction} with the
Q-instanton source and boundary terms and its on-shell action. In
Section \ref{sec:pathintegral} we make a path integral analysis of
the Q-instanton and in Section \ref{sec:Rto4} we discuss the
Q-instanton contribution to the $\mathcal{R}^4$ terms. We end with a
discussion of our results in Section \ref{sec:discussion}.

\section{Q-branes}\label{sec:IIBconjugacyclasses}

Before discussing the Q-instantons we first outline some general
ideas regarding the concept of Q-branes.

The low energy description of the type IIB superstring in a bosonic
background with vanishing 3-form and 5-form field strengths is
described by the well-known axion-dilaton action coupled to gravity
\begin{equation}\label{standardIIBaction}
    S=\int_{\mathcal{M}_{9,1}}\left(\star 1 R-\frac{1}{2}\star d\phi\wedge d\phi-
    \frac{1}{2}e^{2\phi}\star d\chi\wedge d\chi\right)\, ,
\end{equation}

\noindent where the vacuum expectation value of $e^{\phi}$ gives the
string coupling $g_S$. The quantum moduli space of the inequivalent
values of the complex axion--dilaton field $\tau=\chi+ie^{-\phi}$ of
the non-perturbative type IIB string theory is conjectured
\cite{Hull:1994ys} to be given by the orbifold

\begin{equation}\label{quantummodulispace}
    \frac{PSL(2,\mathbb{R})}{SO(2)\times PSL(2,\mathbb{Z})}\,.
\end{equation}

\begin{figure}[h]
\includegraphics[scale=.6]{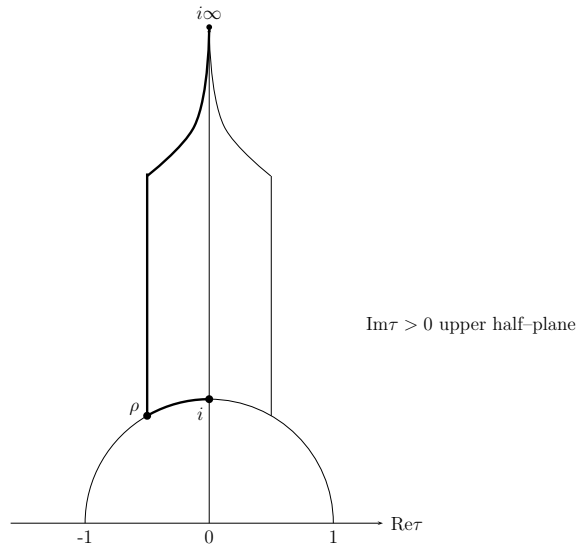}\vspace{-5cm}
\caption{Fundamental domain of $PSL(2,\mathbb{Z})\backslash
PSL(2,\mathbb{R})/SO(2)$}
 \label{fundamentaldomain}
\end{figure}

\noindent This orbifold is depicted in Figure
\ref{fundamentaldomain} \vspace*{-3.5cm}
in which we have exaggerated the cusp-like behavior near the
orbifold point $\tau_0=i\infty$ where the string coupling $g_S$ is
close to zero. By $\rho$ we denote the point $\rho=-{1\over
2}+i{{\sqrt 3}\over{2}}$. It is this region of the moduli space
where we know that there is agreement between results from the
perturbatively defined type IIB superstring theory and IIB
supergravity. For example, the spectrum of the D-branes predicted by
perturbative IIB string theory corresponds to brane--like solutions
of IIB supergravity \cite{Polchinski:1995mt}.

To indicate the place of the new Q7-branes in the fundamental domain
of the axion--dilaton moduli space, let us recall that the
conjectured $SL(2,\mathbb{Z})$ duality of IIB superstring theory
leads to the notion of $(p,q)$ branes, \emph{i.e.} extended objects
on which a $(p,q)$ string ends. The numbers $p$ and $q$ are two
non-negative integers. From the point of view of the fundamental
string, the F1-string or the $(1,0)$ string whose coupling constant
goes to zero in the (perturbative) region of the moduli space at the
point $\tau_0=i\infty$, the $(p,q)$ string is a bound state of $p$
F1-strings and $q$ D1-branes \cite{Witten:1995im}. Alternatively,
one can consider a $(p,q)$ string with $p$ and $q$ relatively prime
as an elementary string whose perturbative sector is determined by
the zero limit of the coupling constant associated with
$e^{\phi_{q,r}}$ \cite{Schwarz:1995dk},
\begin{equation}\label{defT}
    e^{\phi_{q,r}}\equiv qe^{\phi}\mid\tau+\frac{r}{2q}\mid^2\quad\text{with}\quad
    r=2\sqrt{pq}\,,
\end{equation}

\noindent in which case the relevant part of the moduli space is a
cusp around the point $\tau_0=-r/2q$ (where
$e^{\phi_{q,r}}\rightarrow 0$). The low energy description of the
$(p,q)$ string theory is again described by the action
\eqref{standardIIBaction}. This can be made manifest by writing Eq.
\eqref{standardIIBaction} in the following form

\begin{equation}\label{pqIIBaction}
    S=\int_{\mathcal{M}_{9,1}}\left(\star 1 R-\frac{1}{2}\star d\phi_{q,r}\wedge d\phi_{q,r}-
    \frac{1}{2}e^{2\phi_{q,r}}\star d\chi_{q,r}\wedge d\chi_{q,r}\right)\,.
\end{equation}

\noindent The form of the action \eqref{pqIIBaction} suggests that
each $(p,q)$ string vacuum, \emph{i.e.} the point where
$e^{\phi_{q,r}}\rightarrow 0$, has its own coupling constant
$<e^{\phi_{q,r}}>$ and axion $\chi_{q,r}$. The passage from one
$(p,q)$ string vacuum to another and corresponding field
redefinitions are, of course, governed by the $SL(2,\mathbb{Z})$
symmetry of the type IIB string theory.

Let us now introduce, using the notation of
\cite{Bergshoeff:2007aa}, the $SL(2,\mathbb{R})$ algebra valued
charge matrix
\begin{equation}\label{matrixQ}
    Q=\left(
    \begin{array}{cc}
     r/2 & p \\
     -q & -r/2
     \end{array}
    \right)\,,
\end{equation}
which describes the electric coupling of the Q7--brane to an
$SL(2,\mathbb{R})$ triplet of 8--forms (see Eqs.
(\ref{SU(1,1)axion}) and (\ref{A8}) below). The conjugacy classes of
$SL(2,\mathbb{R})$ are characterized by the value of the trace of
$e^Q$,
\begin{equation}\label{eQ}
    e^Q=\cos(\sqrt{\text{det}\,Q})\mathbbm{1}+\frac{\sin(\sqrt{\text{det}\,Q})}{\sqrt{\text{det}\,Q}}Q\,.
\end{equation}
Families of $SL(2,\mathbb{R})$ conjugacy classes are formed by
\begin{equation}
{\rm tr}\,e^Q=2\cos(\sqrt{\text{det}\,Q})\,\left\{
\begin{array}{c}
=2\\
>2\\
<2
\end{array}\right.\quad\text{or equivalently, by}\quad \text{det}\,Q\,\left\{\begin{array}{c}
=0\\
<0\\
>0
\end{array}\right.\,.
\end{equation}

When we add D-branes to the type IIB supergravity theory the duality
group $SL(2,\mathbb{R})$ is broken down to the subgroup that is
generated by the shift symmetry of the RR axion, \emph{i.e.} the
$\mathbb{R}$ subgroup of $SL(2,\mathbb{R})$. This for example
implies that all the D-brane actions are invariant under the shift
of the RR axion. Likewise, when we add a Q-brane to the type IIB
supergravity theory the duality group $SL(2,\mathbb{R})$ is broken
down to the subgroup that is generated by the shift symmetry of the
$\chi'$, \emph{i.e.} the $SO(2)$ subgroup of $SL(2,\mathbb{R})$.
Hence, all the brane solutions of IIB supergravity are associated to
fixed points of $e^Q$ with either $\text{det}\,Q=0$ or
$\text{det}\,Q>0$. \noindent The case $\text{det}\,Q=0$ corresponds
to the $(p,q)$ branes\footnote{ The $SL(2,R)$ covariant actions
which describe the $(p,q)$--branes have been constructed in
\cite{Bergshoeff:2006gs,Bergshoeff:2007ma}. They can be regarded as
the $SL(2,R)$ transformed Dp-brane actions.} and the case
$\text{det}\,Q>0$ corresponds to Q-branes. The case
$\text{det}\,Q<0$ does not arise because there are no fixed points
of $e^Q$ with $\text{det}\,Q<0$ that are part of the quantum moduli
space \eqref{quantummodulispace}. The point $\tau_0$ is a fixed
point under the $e^Q$ transformation if it satisfies the equation

\begin{equation}\label{etoQ}
e^Q\tau_0={{a\tau_0+b}\over{c\tau_0+d}}=\tau_0\hspace{.5cm}\text{where}\hspace{.5cm}
\left(\begin{array}{cc}
a & b \\
c & d
\end{array}
\right)=e^Q\,.
\end{equation}

\noindent The fixed points $\tau_0$ of $e^Q$ with $q>0$ and
$0\le\text{Im}\,\tau_0<\infty$ for $\text{det}\,Q\ge 0$ are given by

\begin{equation}\label{fixed}
\tau_0=-\frac{r}{2q}+\frac{i}{q}\sqrt{\rm{det}\,Q}\,.
\end{equation}

As shown in \cite{Bergshoeff:2007aa} the Q7-brane configurations are
most conveniently described in terms of the variables $T$ and
$\chi'$ which (for $q>0\,${}\footnote{The restriction $q>0$
guarantees that $T$, which is the tension of a Q7-brane, is postive
\cite{Bergshoeff:2007aa}.}) are defined by the following relations

\begin{equation}\label{tautoT}
    \frac{\tau-\tau_0}{\tau-\bar\tau_0}=e^{2i\sqrt{\text{det}\,Q}\,\mathcal{T}}\,,
\end{equation}

\noindent where $\mathcal{T}$ is given by
\begin{equation}\label{mathcalT}
    \mathcal{T}=\chi'+\frac{i}{4\sqrt{\rm{det}\,Q}}\log\frac{T+2\sqrt{\rm{det}\,Q}}{T-2\sqrt{\rm{det}\,Q}}\,,
    \hspace{1cm}\chi'\sim\chi'+\frac{\pi}{\sqrt{\text{det}\,Q}}\,.
\end{equation}

\noindent As follows from Eq. (\ref{tautoT}) the values $\chi'$ and
$\chi'+\pi/\sqrt{\text{det}\,Q}$ are to be identified. In the limit
$\text{det}\,Q\rightarrow 0$ Eq. \eqref{mathcalT} reduces to

\begin{equation}\label{detQ0limit}
    \mathcal{T}\rightarrow\frac{-1}{q(\tau+\frac{r}{2q})}\quad\text{with}\quad
    r=2\sqrt{pq}\,.
\end{equation}

The requirement that $\text{Im}\,\tau>0$ implies that
$T>2\sqrt{\text{det}\,Q}$ or what is the same
$\text{Im}\,\mathcal{T}>0$. The relation \eqref{tautoT} between
$\tau$ and $\mathcal T$ is a conformal mapping from the upper half
plane $\text{Im}\,\tau>0$ to the vertical strip
\begin{equation}\label{classicalmodspace}
\{\,\mathcal{T}\,\mid\,\text{Im}\,\mathcal{T}>0\quad\text{and}\quad
\chi'\sim\chi'+\frac{\pi}{\sqrt{\text{det}\,Q}}\,\}\,.
\end{equation}
For special values of $p,q,$ and $r$ the point $\tau_0$ is equal to
the points $i$ and $\rho$ of Fig. \ref{fundamentaldomain}. In terms
of $\mathcal{T}$ the region close to $i$ or $\rho$ looks like a
cusp, that is, a region where
$\text{Im}\mathcal{T}\rightarrow\infty$ while $\chi'$ becomes
undetermined as $\tau$ approaches $\tau_0$.

Table \ref{tableorbifolddata} shows the values of $p,q,r$ for each
of the orbifold points $\tau_0$ of Fig \ref{fundamentaldomain} and
the related $SL(2,\mathbb{Z})$ conjugacy classes. The periodicity of
the axion $\chi'$ is determined by the value of
$\pi/\sqrt{\text{det}\,Q}$. The discrete isometries are the
$PSL(2,\mathbb{Z})$ transformations generated by $T$ and $S$ with
$T\tau=\tau+1$ and $S\tau=-1/\tau$. The periodicity of $\chi'$
follows from the fact that the transformations $S$ and $T^{-1}S$
are, respectively, of order 2 and 3 in $PSL(2,\mathbb{Z})$.

\begin{table}
\begin{tabular}{|c|c|c|c|c|}
  \hline
  $\tau_0$ & $(p,q,r)$ & $\pi/\sqrt{\text{det}\,Q}$ & discrete isometry & $SL(2,\mathbb{Z})$ conj. class \\
  \hline
  $i\infty$ & $(1,0,0)$ & $\infty$ & $T:\chi\rightarrow\chi+1$ & $T$ \\
  &&&&\\
  $i$ & $(\tfrac{\pi}{2},\tfrac{\pi}{2},0)$ & 2 & $-S:\chi'\rightarrow\chi'+1$ & $-S$\\
  &&&&\\
  $\rho$ & $(\tfrac{2\pi}{3\sqrt{3}},\tfrac{2\pi}{3\sqrt{3}},\tfrac{2\pi}{3\sqrt{3}})$ & 3 & $-T^{-1}S:\chi'
  \rightarrow\chi'+1$ & $-T^{-1}S$\\
  \hline
\end{tabular}
\caption{Properties of the orbifold points
$\tau_0=i\infty,i,\rho$.}\label{tableorbifolddata}
\end{table}

In terms of the fields $T$ and $\chi'$ the action
\eqref{standardIIBaction} takes the form

\begin{equation}\label{pqrIIBaction}
    S=\int_{\mathcal{M}_{9,1}}\left(\star 1 R-\frac{1}{2}\frac{1}{T^2-4\text{det}\,Q}\star dT\wedge dT-\frac{1}{2}(T^2
    -4\text{det}\,Q)\star d\chi'\wedge d\chi'\right)\,.
\end{equation}

\noindent This form of the IIB supergravity action is obtained using
the field redefinition (\ref{tautoT}). The dependence of Eq.
(\ref{pqrIIBaction}) on the parameter $\text{det}\,Q$ can be removed
by making the inverse field redefinition. However, when we couple
the action \eqref{pqrIIBaction} to the Q7-brane action the
dependence on the conjugacy class parameter $\text{det}\,Q$ cannot
be eliminated since there is no field redefinition which takes
\eqref{pqrIIBaction} coupled to a Q7-brane to
\eqref{standardIIBaction} coupled to a D7-brane
\cite{Bergshoeff:2007aa}. The same difference we shall observe in
the case of D- and Q-instantons. This is to be contrasted to the
case of IIB supergravity coupled to $(p,q)$-branes in which case
there always exists a field redefinition that transforms the system
of IIB supergravity coupled to a $(p,q)$-brane to IIB supergravity
coupled to the corresponding D-brane.

Following \cite{Bergshoeff:2007aa} we introduce a 9-form field
strength which is dual to $d\chi'$

\begin{equation}\label{SU(1,1)axion}
    (T^2-4\text{det}Q)\,d\chi'=\star (pF_9+qH_9+rG_9)\equiv\star\mathcal{F}_9\,,
\end{equation}

\noindent where the 9-forms $F_{9},H_9,G_9$ are organized in a
triplet transforming in the adjoint of $SL(2,\mathbb{R})$ and $p,q$
and $r$ are the components of the matrix $Q$ (\ref{matrixQ}). From
the axion $\chi'$ equation of motion (when ignoring its coupling to
the 2-forms and 6-forms) it follows that
\begin{equation}\label{BI}
d\mathcal{F}_9=0\,,
\end{equation}
so that locally
\begin{equation}\label{dA8}
\mathcal{F}_9=d\mathcal{A}_8.
\end{equation}
 The Q7--brane
minimally couples to $\mathcal{A}_8$ via the Wess--Zumino term
\begin{equation}\label{A8}
S^{Q7}_{min}=m\int\,{\mathcal A}_8\,,
\end{equation}
where $m$ is the Q7-brane electric charge with respect to ${\mathcal
A}_8$, or magnetic charge associated with its axion dual $\chi'$. In
\cite{Gibbons:1995vg} it has been shown that the axion charge $m$ of
the $(p,q)$ 7-branes takes discrete values $(m=1,2,3,\cdots)$ while
the discreteness of $m$ for the Q7-branes was shown in
\cite{Bergshoeff:2007aa}.

When $p=1$ and $q=r=0$ Eq. \eqref{SU(1,1)axion} describes the
duality between the RR axion $\chi$ and the RR 8-form $C_8$ whose
field strength is $F_9$. When $q=1$ and $p=r=0$ the field strength
is $H_9=dB_8$ where $B_8$ is an NSNS 8-form that couples to the NSNS
7-brane (the S-dual transformed D7-brane). The case $p=q=0$ and
$r\neq 0$ corresponds to $\text{det}\,Q<0$ and hence there is no
7-brane that couples only to a $D_8$ field whose field strength is
$G_9=dD_8$.

The dynamics of the 9-form $\mathcal{F}_9$ can be described by the
following first order action

\begin{align}
    S\,[g_{\mu\nu}, \chi', \mathcal{F}_9, T]=&\int_{{\mathcal{M}}_{9,1}}\left(*1 R-\frac{1}{2}\frac{1}{T^2-4\text{det}\,Q}
    \star dT\wedge dT-\right.\nonumber\\
    &\left.\frac{1}{2}\frac{1}{T^2-4\text{det}\,Q}\star \mathcal{F}_{9}\wedge\mathcal{F}_9
    -\chi' d\mathcal{F}_{9}\right)\,.\label{firstorderchi'}
\end{align}

\noindent In the action (\ref{firstorderchi'}) the axion $\chi'$
appears (in a shift symmetry invariant way) as a Lagrange
multiplier. The term $d\mathcal{F}_9$ is parity odd as is
$\chi'\,$\footnote{The parity oddness of $\chi'$ can be understood
as follows. The field redefinition \eqref{tautoT} and
\eqref{mathcalT} implies that we have
\begin{equation}\label{relationforparity}
-e^{\phi}(\chi-\text{Re}\,\tau_0)=\frac{(T^2-4\text{det}\,Q)^{1/2}}{2\sqrt{\text{det}\,Q}}\,\sin
2\sqrt{\text{det}\,Q}\,\chi'\,.
\end{equation}
Then it follows from the relation \eqref{relationforparity} that
$\chi'$ has the same parity as $\chi$. Since, the RR axion is parity
odd so is $\chi'$.}. The variation of \eqref{firstorderchi'} with
respect to $\mathcal{F}_9$ gives the duality relation
\eqref{SU(1,1)axion}. If we substitute this relation back into the
action we obtain the action \eqref{pqrIIBaction}. If we vary
\eqref{firstorderchi'} with respect to $\chi'$ we find the Bianchi
identity for $\mathcal{F}_9$ (\ref{BI}). If we substitute its
solution (\ref{dA8}) back into the action \eqref{firstorderchi'} we
obtain a second order action for $\mathcal{A}_8$. The action
\eqref{firstorderchi'} will be the starting point of our discussion
of the Q-instantons.

\section{D--instantons}\label{sec:Dinstanton}
Before discussing the new Q-instanton solutions of IIB supergravity,
let us briefly review the derivation of the D-instanton solution
\cite{Gibbons:1995vg}. This is a solution of the equations of motion
of the axion and dilaton coupled to gravity in Euclidean space. The
Wick rotation of the action (\ref{standardIIBaction}) is carried out
by taking into account that the axion is an axial scalar and hence
gets replaced with $i\chi\,$\footnote{We anticipate that in Section
\ref{sec:pathintegral} it will be shown that from the path integral
point of view it is not allowed to send $\chi$ to $i\chi$ (or
$\chi'$ to $i\chi'$ when it concerns the Q-instanton) under a Wick
rotation. Since, in this and the next Section we discuss classical
Euclidean field theory which only provides on-shell information
about the saddle point approximation there is no harm done in
sending $\chi$ to $i\chi$.}. The Wick rotation thus changes the sign
of the Einstein term and the dilaton kinetic term leaving intact the
sign of the axion kinetic term. So the Euclidean action is
\begin{equation}\label{standardE}
    S=\int_{\mathcal{M}_{10}}\left(-\star 1
    R+\frac{1}{2}\star d\phi\wedge d\phi-\frac{1}{2}e^{2\phi}\star d\chi\wedge
    d\chi\right)\,.
\end{equation}
Note that the action is invariant under the axion shift symmetry
$\chi\,\rightarrow \, \chi+ b$ with $b$ being a constant real
parameter. The Einstein equations and the equations of motion of the
axion and the dilaton, which follow from (\ref{standardE}), have the
form
\begin{equation}\label{R}
R_{mn}-{1\over 2}\,(\partial_m\phi\partial_n
\phi-e^{2\phi}\partial_m\chi\partial_n\chi)=0\,,
\end{equation}
\begin{equation}\label{chi}
D_m(e^{2\phi}D^m\chi)=0\,,
\end{equation}
\begin{equation}\label{phi}
D_mD^m\,\phi+e^{2\phi}\,(\partial\chi)^2=0\,.
\end{equation}

The Ansatz imposed on the fields to get the D-instanton solution of
(\ref{R})--(\ref{phi}) is
\begin{equation}\label{ansatz}
g_{mn}=\delta_{mn},\quad d\chi=\pm e^{-\phi}\,d\phi=\mp de^{-\phi}.
\end{equation}
The equation (\ref{ansatz}) is nothing but the Bogomol'nyi bound
saturation condition imposed on the axion--dilaton system in flat
space. The upper and lower signs in (\ref{ansatz}) correspond,
respectively to the D-instanton and anti--D-instanton. When
(\ref{ansatz}) is imposed Eqs. (\ref{R})--(\ref{phi}) reduce to
\begin{equation}\label{chi1}
\partial_m(e^{2\phi}\partial^m\chi)=0\,\quad \rightarrow \quad \quad
\partial^2\,e^\phi=0\,,
\end{equation}
\begin{equation}\label{phi1}
\partial_m\partial^m\phi+(\partial\phi)^2=0 \quad \rightarrow
\quad e^{-\phi}\,\partial^2\,e^\phi=0.
\end{equation}

A spherically symmetric solution to the above equations which
describes a single \linebreak 
(anti-)instanton is
\begin{equation}\label{is}
e^\phi=e^{\phi_\infty}+\frac{c}{r^8}, \quad
\chi-\chi_\infty=\mp(e^{-\phi}-e^{\phi_\infty})\,,
\end{equation}
where the upper sign stands for the instanton and the lower sign
corresponds to the anti--instanton, $\phi_\infty$ and $\chi_\infty$
are the values of the dilaton and axion at $r=\sqrt{x^mx_m}=\infty$
and $c>0$ is (roughly speaking) the instanton charge, namely,
\begin{equation}\label{parameterc}
    c=\frac{2\pi\vert n\vert}{8\text{Vol}(S^9)}\,,
\end{equation}

\noindent with $\text{Vol}(S^9)$ being the volume of a 9-sphere of a
unit radius and $n$ being an integer which manifests the instanton
charge quantization \cite{Gibbons:1995vg}. Note that from (\ref{is})
it follows that for the instanton $\chi+ e^{-\phi}$ is constant and
for the anti-instanton $\chi- e^{-\phi}$ is constant everywhere in
10d space.

The solution (\ref{is}) is singular at $r=0$ which implies that it
is sourced by a point-like object (the instanton) sitting at $r=0$.
The (anti-)instanton contribution to the right hand side of the
axion--dilaton field equations (\ref{chi1}) and (\ref{phi1}) is as
follows
\begin{equation}\label{delta}
\partial_m(e^{2\phi}\partial^m\chi)=\mp 2\pi\vert n\vert\,\delta ^{(10)}(\vec x)\,,\qquad
e^{-\phi}\,\partial^2\,e^\phi=-2\pi\vert n\vert\,e^{-\phi}\,\delta
^{(10)}(\vec x)\,.
\end{equation}
Eqs. (\ref{delta}) can be obtained by varying the supergravity
action (\ref{standardE}) coupled to the instanton source
\begin{equation}\label{standardE+i}
    S=\int_{\mathcal{M}_{10}}\left(-\star 1
    R+\frac{1}{2}\star d\phi\wedge d\phi-\frac{1}{2}e^{2\phi}\star d\chi\wedge
    d\chi\right)+2\pi\vert n\vert\,\,\int_{{\mathcal{M}}_{10}}\delta^{(10)}(\vec x)
    \left(e^{-\phi}\pm\chi\right)\star 1\,,
\end{equation}
and imposing the Ansatz (\ref{ansatz}).

The presence of the instanton source term breaks the invariance of
the action (\ref{standardE+i}) under the shift symmetry $\chi
\,\rightarrow\,\chi+b$. The invariance can be restored by adding to
Eq. (\ref{standardE+i}) the boundary term
\begin{equation}\label{boundary}
-\int_{\partial\mathcal{M}_{10}}\,\chi\, e^{2\phi}\,\star
\,d\chi=-\int_{\mathcal{M}_{10}}\,d(\chi\, e^{2\phi}\,\star
\,d\chi)=\int_{\mathcal{M}_{10}}\,d^{10}x\,\partial_m(\chi\, e^{2\phi}\,
\,\partial^m\chi)\,,
\end{equation}
such that
\begin{equation}\label{boundary1}
\int_{\partial\mathcal{M}_{10}}\, e^{2\phi}\,\star
\,d\chi=\pm\,2\pi\vert n\vert\,\,\int_{{\mathcal{M}}_{10}}\delta^{(10)}(\vec x)\star 1
=\pm\,2\pi\vert n\vert\,.
\end{equation}
Note that this boundary condition is compatible with Eqs.
(\ref{delta}).

The appearance of the boundary term (\ref{boundary}) in the
supergravity action can be best understood if one starts from the
action which includes the field strength $F_9=dA_8$ of the 8--form
gauge field $A_8$ and then dualizes it into the axion action by
adding the term $\int_{\mathcal{M}_{10}}\,\chi d F_9$ (compare with
(\ref{firstorderchi'}))
\begin{equation}\label{action9}
S=\int_{\mathcal{M}_{10}}\left(-\star 1
    R+\frac{1}{2}\star d\phi\wedge d\phi+\frac{1}{2}e^{-2\phi}\star F_9\wedge
    F_9\right)+\int_{\mathcal{M}_{10}}\,\chi\, dF_9\,.
\end{equation}
If in (\ref{action9}) the field $F_9$ is considered as the
independent one (\emph{i.e.} not a curl of $A_8$), the variation
with respect to this field gives the duality relation
$F_9=e^{2\phi}\,\star d\chi$ which can be substituted back into the
action (\ref{action9}) thus reducing it to
\begin{equation}\label{standardE+s}
    S=\int_{\mathcal{M}_{10}}\left(-\star 1
    R+\frac{1}{2}\star d\phi\wedge d\phi-\frac{1}{2}e^{2\phi}\star d\chi\wedge
    d\chi\right)-\int_{\partial\mathcal{M}_{10}}\,\chi\, e^{2\phi}\,\star
\,d\chi\,.
\end{equation}
The boundary term we have looked for has appeared as a result of the
integration by parts of the last term in (\ref{action9}). To
summarize, the shift symmetry invariant action for the IIB
supergravity - D--instanton system is
\begin{eqnarray}\label{standardE+s+i}
    S&=&\int_{\mathcal{M}_{10}}\left(-\star 1
    R+\frac{1}{2}\star d\phi\wedge d\phi-\frac{1}{2}e^{2\phi}\star d\chi\wedge
    d\chi\right)-\int_{\partial\mathcal{M}_{10}}\,\chi\, e^{2\phi}\,\star
\,d\chi\nonumber\\
&&\\
&&+2\pi\vert n\vert\,\,\int_{{\mathcal{M}}_{10}}\delta^{(10)}(\vec
x)
    \left(e^{-\phi}\pm\chi\right)\star 1\,.\nonumber
\end{eqnarray}

We are now ready to compute the on--shell value of this action by
substituting into (\ref{standardE+s+i}) the instanton solution
(\ref{ansatz}), (\ref{is}). Then the bulk part of the action
vanishes because of the Bogomol'nyi bound saturation, the
contribution from the boundary term gets canceled by the $\chi$ part
of the source term, and we are left with
\begin{equation}\label{onshellD}
S_D|_{\rm on-shell}=2\pi\vert n\vert\, e^{-\phi_\infty},
\end{equation}
where $e^{\phi_\infty}$ is the string coupling constant.

In Section \ref{sec:pathintegral} it will be shown that the result
\eqref{onshellD} corresponds to a saddle point approximation of a
path integral that computes the transition amplitude between axion
conjugate momentum eigenstates or, what is the same, between Noether
charge eigenstates of the Noether current, associated to the shift
symmetry $\chi\rightarrow\chi+b$, that differ by $n$ units. As shown
in Section \ref{sec:pathintegral} (see also \cite{Green:1997tv}), in
order to obtain a saddle point approximation between axion $\chi$
eigenstates, one must add to \eqref{onshellD} the imaginary term
\begin{equation}\label{imchi}
 -2\pi ni\chi_{\infty}\,,
\end{equation}

\noindent with $n>0$ for the D-instanton and $n<0$ for the
anti-D-instanton. The axion that appears in \eqref{imchi} is the RR
axion $\chi$ of the Lorentzian IIB theory (and not the Wick rotated
one of this Section). Thus the D-instanton action takes the form
\begin{equation}\label{onshelld1}
  S_D= -2\pi i |n|\tau_{\infty}\,.
\end{equation}

\section{Q-instantons}\label{sec:Euclideanaction}

Let us now perform an analysis similar to the one described above to
find instanton solutions of IIB supergravity for which
$\text{det}\,Q>0$ using the fields $(T,\chi')$.

\subsection{Q-instanton action}
The analog of Eq. (\ref{action9}), that should provide us with the
relevant boundary term in the Q-instanton action, is the Euclidean
version of the action \eqref{firstorderchi'}, namely
\begin{equation}\label{Euclideanized}
    S=\int_{{\mathcal{M}}_{10}}\left(-\star 1
    R+\frac{1}{2}\frac{1}{T^2-4\text{det}\,Q}\star dT\wedge dT+
    \frac{1}{2}\frac{1}{T^2-4\text{det}\,Q} \star {\mathcal F}_{9}\wedge
    {\mathcal F}_{9}+\chi'
d{\mathcal F}_{9}\right)\,,
\end{equation}

\noindent where we replaced $\chi'$ by $i\chi'$. The $\mathcal{F}_9$
equation of motion gives the duality relation between
$\mathcal{F}_9$ and the Wick rotated $\chi'$ (similar to
(\ref{SU(1,1)axion})). Substituting the duality relation back into
the action we get
\begin{equation}\label{Euclidean second order axion II}
\begin{aligned}
    S= &\int_{{\mathcal{M}}_{10}}\left(-\star 1
    R+\frac{1}{2}\frac{1}{T^2-4\text{det}\,Q}\,\star dT\wedge dT-
    \frac{1}{2}(T^2-4\text{det}\,Q) \,\star d\chi' \wedge d\chi'\right)\\ &-\int_{{\mathcal{\partial{M}}}_{10}}
    (T^2-4\text{det}\,Q)\,\chi'
    \star d\chi'\,.
\end{aligned}
\end{equation}

To this action we should couple an instanton source term that will
be the counterpart of (\ref{standardE+s+i}) in the $(T,\chi')$
basis. Remember that the form of the D-instanton coupling term was
prompted by the structure of the source terms on the right-hand side
of the axion--dilaton Eqs. (\ref{delta}) which take care of the
singularity of the D-instanton solution (\ref{is}). So to find the
relevant form of the instanton coupling term in the new basis we
should study the $(T,\chi')$ equations of motion.

As in the D-instanton case we shall assume that for the solution
under consideration the $d=10$ space is flat and, as follows from
the Einstein equation, $T$ and $\chi'$ are related by the following
Bogomol'nyi bound saturation condition
\begin{equation}\label{ansatz1}
d\chi'=\pm (T^2-4\text{det}\,Q)^{-1}\,dT\,.
\end{equation}
Then the $\chi'$-- and T--field equations that follow from
(\ref{Euclidean second order axion II}) and that satisfy
\eqref{ansatz1} take, respectively, the following form
\begin{equation}\label{axion'}
-\partial_m((T^2-4\text{det}\,Q)\,\partial^m\,\chi')=0,
\end{equation}
\begin{equation}\label{T}
(T^2-4\text{det}\,Q)^{-1}\,\partial_m\partial^m
T=\frac{1}{4\sqrt{\text{det}\,Q}}
\left(\log\frac{T+2\sqrt{\text{det}\,Q}}{T-2\sqrt{\text{det}\,Q}}\right)'\,\partial_m\partial^m
T=0\,,
\end{equation}
where the prime over the logarithm denotes its derivative with
respect to $T$. From the form of (\ref{axion'}) and (\ref{T}) we see
that for these equations to acquire the delta--function source terms
$2\pi \vert n\vert \,\delta^{(10)}(\vec x)$ we should add to the
action (\ref{Euclidean second order axion II}) a coupling term of
the form\footnote{The source term \eqref{sourcenewbasis} is uniquely
specified by requiring an electric coupling term linear in $\chi'$
and by requiring it to preserve the same supersymmetries as the
D-instanton.}
\begin{equation}\label{sourcenewbasis}
   2\pi \vert n\vert \,\int_{{\mathcal{M}}_{10}}\delta^{(10)}(\vec x)
    \left(\frac{1}{4\sqrt{\text{det}\,Q}}
    \log\frac{T+2\sqrt{\text{det}\,Q}}{T-2\sqrt{\text{det}\,Q}} \pm\chi'\right)\star
    1\,.
\end{equation}

\noindent The source term guarantees that the instanton solution we
are interested in is defined on the entire 10d Euclidean space. As
in the D-instanton case, the Q-instanton charge is quantized
$(|n|=1,2,3,\cdots)$ as we shall demonstrate in the next Subsection.

Upon adding (\ref{sourcenewbasis}) to Eq. (\ref{Euclidean second
order axion II}) we get the following action in flat Euclidean space
\begin{equation}\label{Euclidean second order axion III}
\begin{aligned}
    S= &\int_{{\mathcal{M}}_{10}}\left(\frac{1}{2}\frac{1}{T^2-4\text{det}\,Q}dT\wedge
    \star dT-
    \frac{1}{2}(T^2-4\text{det}\,Q) d\chi' \wedge\star d\chi'\right)\\
     &-\int_{{\mathcal{\partial{M}}}_{10}}
    (T^2-4\text{det}\,Q)\,\chi'
    \star d\chi'\\& +2\pi|n|\,\int_{{\mathcal{M}}_{10}}\delta^{(10)}(\vec x)
    \left(\frac{1}{4\sqrt{\text{det}\,Q}}\log\frac{T+2\sqrt{\text{det}\,Q}}
    {T-2\sqrt{\text{det}\,Q}}\pm
    \chi'\right)\star 1\,.
\end{aligned}
\end{equation}

\noindent As in the D-instanton case, due to the presence of the
source term there is only one boundary, $\partial \mathcal{M}_{10}$,
which is located at $r=\sqrt{x^mx_m}=\infty$. The action
\eqref{Euclidean second order axion II} is invariant under arbitrary
shifts of the axion, $\chi'\rightarrow\chi'+b$ (where $b$ is any
real number) provided that
\begin{equation}\label{boundary2}
\int_{{\mathcal{\partial{M}}}_{10}}
    \,(T^2-4\text{det}\,Q)
    \star d\chi'=\pm\,2\pi|n| \,\int_{{\mathcal{M}}_{10}}\,\delta^{(10)}(\vec x)\,\star
    1\,=\pm\,2\pi|n|.
\end{equation}
The equations of motion of the fields $\chi'$ and $T$ which follow
from Eq. \eqref{Euclidean second order axion III} acquire the
contribution of the instanton source term and take the following
form on the Bogomol'nyi bound (\ref{ansatz1})
\begin{equation}\label{axion'1}
\partial_m((T^2-4\text{det}\,Q)\,\partial^m\,\chi')=\mp\,2\pi|n|\delta^{(10)}(\vec
x)\,,
\end{equation}
\begin{equation}\label{T1}
\partial_m\partial^m T=-2\pi|n|\delta^{(10)}(\vec x)\,.
\end{equation}

The Q-instanton solution to Eqs. (\ref{ansatz1}) and (\ref{T1}) is
\begin{align}\label{Qsolution}
    & \chi'-\chi'_{\infty}=\mp\left[
    \frac{1}{4\sqrt{\text{det}\,Q}}\log\left(\frac{T+2\sqrt{\text{det}\,Q}}
    {T-2\sqrt{\text{det}\,Q}}\right)-\frac{1}{4\sqrt{\text{det}\,Q}}\log\left(\frac{T_{\infty}+2\sqrt{\text{det}\,Q}}
    {T_{\infty}-2\sqrt{\text{det}\,Q}}\right)\right]\,,\\
    & T=T_{\infty}+\frac{c}{r^8}\,,
\end{align}

\noindent where $c$ has been given in Eq. \eqref{parameterc}.
Substituting this solution into the action \eqref{Euclidean second
order axion III} we get the on--shell value of the Q-instanton
action
\begin{equation}\label{onshellactionQ}
    S_Q\mid_{\text{on-shell}}=\frac{\pi\vert
    n\vert }{2\sqrt{\rm{det}\,Q}}\log\frac{T_\infty+2\sqrt{\rm{det}\,Q}}
    {T_\infty-2\sqrt{\rm{det}\,Q}}\,.
\end{equation}
By virtue of the definition of $T$, Eqs. (\ref{tautoT}) and
(\ref{mathcalT}), when $\rm{det}\,Q\rightarrow 0$ the action
(\ref{onshellactionQ}) reduces to the $(p,q)$-instanton action where
$p$ and $q$ are as in Eq. \eqref{detQ0limit}.

In Section \ref{sec:pathintegral} it will be shown that the result
\eqref{onshellactionQ} corresponds to a saddle point approximation
of a path integral that computes the transition amplitude between
Noether charge eigenstates of the Noether current, associated to the
shift symmetry $\chi'\rightarrow\chi'+b$, that differ by $n$ units.
As will also be shown in Section \ref{sec:pathintegral}, in order to
obtain a saddle point approximation between axion $\chi'$
eigenstates one must add to \eqref{onshellD} the imaginary term
\begin{equation}\label{imagpart}
    -2\pi ni\chi'_{\infty}\,,
\end{equation}

\noindent with $n>0$ for a Q-instanton and $n<0$ for an
anti-Q-instanton. The axion that appears in \eqref{imagpart} is the
axion $\chi'$ of the Lorentzian IIB theory (and not the Wick rotated
one of this Section). The Q-instanton action thus acquires the form
\begin{equation}\label{qi}
S_Q=-2\pi i |n|{\mathcal T}_{\infty}\,.
\end{equation}

It is instructive to compare the actions \eqref{Euclidean second
order axion III} and (\ref{onshellactionQ}) and the equations of
motion (\ref{axion'1}) and (\ref{T1}) with the D-instanton case of
the previous Section. This will allow us to understand at which
point the Q-instanton solutions to \eqref{Euclidean second order
axion III} are different from the D-instanton solutions. To this end
let us perform the following field redefinition\footnote{The field
redefinition \eqref{field transformations} does not follow from the
Lorentzian IIB field redefinitions \eqref{tautoT} and
\eqref{mathcalT} via Wick rotation. Here we consider the way in
which the D- and Q-instantons differ from the point of view of
classical Euclidean field theory.\label{noncommutingtrafos}} whose
form is prompted by the fact that the Bogomol'nyi bounds
(\ref{ansatz}) and (\ref{ansatz1}) must be field redefinition
equivalent as both follow from the Einstein equation with
$g_{\mu\nu}=\delta_{\mu\nu}$. The field redefinition relating
\eqref{ansatz}
 and \eqref{ansatz1} is
\begin{equation}\label{field transformations}
T=2\sqrt{\text{det}\,Q} \, e^{\phi} \chi\,, \hspace{.5cm}
\chi'=\frac{1}{4\sqrt{\text{det}\,Q}}\log\left(\chi^2-e^{-2\phi}\right)\,.
\end{equation}
Using \eqref{field transformations} the action \eqref{Euclidean
second order axion II} in flat space takes the form
\begin{eqnarray}\label{oldbasis}
    S&=&\int_{\mathcal{M}_{10}}\left(\frac{1}{2}\star d\phi\wedge d\phi-\frac{1}{2}e^{2\phi}\star d\chi\wedge
    d\chi\right)\\
   &&\hspace{-30pt} -\frac{1}{2}\,
   \int_{\partial\mathcal{M}_{10}}\,\log\left(\chi^2-e^{-2\phi}\right)\, \star
\,(e^{2\phi}\,\chi\,d\chi+d\phi)\pm\frac{\,2\pi|n|}{2\sqrt{\text{det}\,Q}}\,
\int_{{\mathcal{M}}_{10}}
\delta^{(10)}(\vec
x)\,
     \log(\chi\pm e^{-\phi})\star 1\,.\nonumber
\end{eqnarray}
We observe that the bulk part of (\ref{oldbasis}) coincides with the
bulk part of the action (\ref{standardE+s+i}) while the boundary and
source terms of \eqref{oldbasis} and \eqref{standardE+s+i} differ.
This is in agreement with the remark made in Section
\ref{sec:IIBconjugacyclasses} regarding field redefinitions and
Q-branes: there exists no field redefinition that relates IIB
supergravity coupled to a Q-brane to IIB supergravity coupled to a
D-brane.

Consider the equations of motion of $\chi$ and $\phi$ that follow
from (\ref{oldbasis}). These are
\begin{equation}\label{chi2}
\partial_m(e^{2\phi}\partial^m\chi)=\mp\frac{\pi |n|}{\sqrt{\text{det}\,Q}\,
(e^{-\phi}\pm\chi)}\delta^{(10)}(\vec x),
\end{equation}
\begin{equation}\label{phi2}
\partial_m\partial^m\,\phi+e^{2\phi}\,(\partial\chi)^2=-\frac{\pi|n|\,
e^{-\phi}}{\sqrt{\text{det}\,Q}\,(e^{-\phi}\pm
\chi)}\delta^{(10)}(\vec x).
\end{equation}
Using the Bogomol'nyi bound (\ref{ansatz}) which is related to
\eqref{ansatz1} via Eqs. \eqref{field transformations} one reduces
Eqs. \eqref{chi2} and (\ref{phi2}) to
\begin{equation}\label{chiphi}
\partial_m(e^{2\phi}\partial^m\chi)=\mp\frac{\pi|n|}{\sqrt{\text{det}\,Q}\,(e^{-\phi}\pm \chi)}\delta^{(10)}(\vec
x)\quad \rightarrow\quad
\partial_m\partial^m(e^{\phi})=-\frac{\pi|n|}{\sqrt{\text{det}\,Q}\,(e^{-\phi}\pm\chi)}\delta^{(10)}(\vec
x)\,.
\end{equation}
Comparing (\ref{delta}) with (\ref{chiphi}) we see that the
difference is in the factor
$1\over{2\sqrt{\text{det}\,Q}\,(e^{-\phi}\pm\chi)}$ in the source
term of the latter.

The Ansatz (\ref{ansatz}) and Eqs. (\ref{chiphi}) must be compatible
with the boundary condition (\ref{boundary2}) required by the shift
symmetry of the axion $\chi'$. Indeed, in terms of $\phi$ and $\chi$
which satisfy (\ref{ansatz}), Eq. (\ref{boundary2}) takes the form
\begin{equation}\label{boundary2old}
\int_{\partial\mathcal{M}_{10}}\,2\,\sqrt{\text{det}\,Q}\,(e^{-\phi}\pm\chi)\, e^{2\phi}\,\star
\,d\chi
=\mp\,{2\pi|n|}\,
\int_{{\mathcal{M}}_{10}}\delta^{(10)}(\vec x)\star 1\,,
\end{equation}
which is consistent with (\ref{chi2}) and (\ref{chiphi}).

Finally, in terms of the boundary values of $\phi$ and $\chi$ the
on-shell action (\ref{onshellactionQ}) for the Q-instanton has the
form
\begin{equation}\label{onshellactionQ1}
    S_Q\mid_{\text{on-shell}}=\frac{\pi|n|}{2\sqrt{\rm{det}\,Q}}\,
    \log\frac{\chi_\infty+e^{-\phi_\infty}}
    {\chi_\infty-e^{-\phi_\infty}}\,,
\end{equation}
which obviously differs from the D-instanton on--shell action
(\ref{onshellD}). We conclude that from the point of view of the
classical Euclidean field theory the difference between the D- and
the Q-instanton lies in the different source and boundary terms.

In \eqref{onshellactionQ1} $\chi_{\infty}$ and $\phi_{\infty}$
appear in the combinations $\chi_{\infty}+e^{-\phi_{\infty}}$ and
$\chi_{\infty}-e^{-\phi_{\infty}}$, that are naturally associated to
the coset $SL(2,\mathbb{R})/SO(1,1)$, and not to the coset
$SL(2,\mathbb{R})/SO(2)$. The Euclidean path integral is invariant
under $SL(2,\mathbb{R})/SO(2)$ nonlinear transformations (see
Section \ref{sec:pathintegral}). It is therefore not possible to
obtain \eqref{onshellactionQ1} with $\chi_{\infty}$ and
$\phi_{\infty}$ denoting the RR axion and the dilaton, respectively,
from a path integral analysis. This shows from a somewhat different
point of view that the Q-instanton differs from the D-instanton.

As will be shown in Section \ref{sec:pathintegral} the starting
point of the path integral analysis is the first order action
\eqref{firstorderchi'}. From the path integral perspective the
distinction between a Q- and a D-instanton is that there does not
exist a local field redefinition that relates the action
\eqref{firstorderchi'} for $\text{det}\,Q=0$ to the action
\eqref{firstorderchi'} for $\text{det}\,Q>0$. The reason being that
\eqref{firstorderchi'} depends both on $\mathcal{F}_9$ and $\chi'$.

\subsection{Q-instanton charge quantization}

The quantization of the Q-instanton charge, Eq. \eqref{boundary2},
follows from the standard Dirac--\linebreak Nepomechie--Teitelboim
quantization condition
\cite{Dirac:1948um,Nepomechie:1984wu,Teitelboim:1985yc} applied to
the Q(-1)-brane (instanton) and a Euclidean Q7-brane in a way
similar to the D-instanton case \cite{Gibbons:1995vg}. Assume that
the spatial volume of the 7-brane is compact with the topology of
$S^7$. If we keep one point on the $S^7$ surface fixed and transport
the 7-brane along closed paths its world-volume will have the
topology of $S^8$. The wave function of this compact 7-brane will
acquire, due to its minimal coupling to the axion dual 8--form
(\ref{A8}) (with $m=1$), the following phase factor

\begin{equation}\label{wf}
    e^{i\int_{\Sigma} {\mathcal A}_8}\,,
\end{equation}

\noindent where $\Sigma$ is the world-volume of the compact 7-brane.
Using Stokes' theorem we can write

\begin{equation}
    \int_{\Sigma}{\mathcal A}_8=\int_{S}{\mathcal F}_9=-\int_{S'}{\mathcal F}_9\,,
\end{equation}

\noindent where $S$ and $S'$ are the two capping surfaces of the
world-volume $\Sigma=S^8$. The single-valuedness of the wave
function (\ref{wf}) requires that

\begin{equation}\label{diracquant}
    \int_{S^9}{\mathcal F}_9=2\pi
    n\,,
\end{equation}

\noindent where $S^9=S\bigcup S'$. Taking now into account the
duality relation between $\mathcal{F}_9$ and the Wick rotated axion
$\chi'$, as follows by varying Eq. \eqref{Euclideanized} with
respect to $\mathcal{F}_9$, we arrive at Eq. (\ref{boundary2}) which
relates the value of the Q--instanton boundary term to its
\emph{quantized} charge.

\subsection{The half-BPS condition}\label{Qsusy}

We will show that the Bogomol'nyi bound \eqref{ansatz1} also follows
by analyzing the Killing spinor equations. In the Lorentzian IIB
theory with vanishing 3- and 5-form field strengths the Killing
spinor equations are

\begin{align}
    &\delta\Psi_{m}=(\nabla_{m}-\frac{i}{2}Q_{m})\epsilon\,,\label{deltaPsi}\\
    &\delta\lambda=iP_{m}\gamma^{m}\epsilon_C\,,\label{deltalambda}
\end{align}

\noindent where $\epsilon=\epsilon_1+i\epsilon_2$ and
$\epsilon_C=\epsilon_1-i\epsilon_2$ with $\epsilon_1$ and
$\epsilon_2$ being Majorana--Weyl spinors. The coset Zweibein
$P_{m}$ and the composite $U(1)$ connection $Q_{m}$ of the
axion-dilaton coset space ${{SL(2,{\mathbb R})}\over{SO(2)}}\sim
{{SU(1,1)}\over{SU(1)}}$ must satisfy the Bianchi identity

\begin{equation}
    dP-2iQ\wedge P=0\,,
\end{equation}

\noindent where $P_{m}$ is such that the action \eqref{pqrIIBaction}
can be written as follows (see \emph{e.g.}
\cite{Schwarz:1983qr,Bergshoeff:2007aa} for details)
\begin{equation}
    S=\int_{\mathcal{M}_{9,1}}\left(\star 1R-2\star P\wedge \bar P\right)\,.
\end{equation}

\noindent We choose a $U(1)$ gauge in which

\begin{align}
    & P_{m}=\frac{1}{2}\frac{\partial_{m}T}{(T^2-4\text{det}\,Q)^{1/2}}+
    \frac{i}{2}(T^2-4\text{det}\,Q)^{1/2}\partial_{m}\chi'\,,\\
    & Q_{m}=\frac{T}{2}\partial_{m}\chi'\,.
\end{align}

We Wick rotate Eqs. \eqref{deltaPsi} and \eqref{deltalambda} by
sending $\chi'$ to $i\chi'$. Treating Eqs. \eqref{deltaPsi} and
\eqref{deltalambda} and their complex conjugates separately we
obtain

\begin{align}
    & \left(\nabla_{m}+\frac{T}{4}\partial_{m}\chi'\right)\epsilon=0\,,\\
    & \left(\nabla_{m}-\frac{T}{4}\partial_{m}\chi'\right)\epsilon_C=0\,,\\
    & \left(\frac{\partial_{m}T}{(T^2-4\text{det}\,Q)^{1/2}}-
    (T^2-4\text{det}\,Q)^{1/2}\partial_{m}\chi'\right)\gamma^{m}\epsilon_C=0\,,\\
    & \left(\frac{\partial_{m}T}{(T^2-4\text{det}\,Q)^{1/2}}+
    (T^2-4\text{det}\,Q)^{1/2}\partial_{m}\chi'\right)\gamma^{m}\epsilon=0\,,
\end{align}

\noindent where $\epsilon$ and $\epsilon_C$ are Wick rotated
spinors. The 1/2 BPS condition for the Q-instanton is
\begin{equation}\label{halfBPSQ}
    \partial_m\chi'=(T^2-4\text{det}\,Q)^{-1}\partial_m T\,,\qquad\epsilon=0\,,
\end{equation}

\noindent and for the anti-Q-instanton

\begin{equation}\label{halfBPSantiQ}
    \partial_m\chi'=-(T^2-4\text{det}\,Q)^{-1}\partial_m T\,,\qquad\epsilon_C=0\,.
\end{equation}

\noindent When either \eqref{halfBPSQ} or \eqref{halfBPSantiQ} holds
we have that the Q- or anti-Q-instanton source term
\eqref{sourcenewbasis} is half BPS. Using $g_{mn}=\delta_{mn}$ it
follows that for the anti-Q-instanton the Killing spinor $\epsilon$
is given by

\begin{equation}
    \epsilon=(T^2-4\text{det}\,Q)^{1/8}\epsilon_{0}\,,
\end{equation}

\noindent where $\epsilon_{0}$ is a constant spinor.

\section{Path integral approach to Q-instantons}\label{sec:pathintegral}

In this Section we will justify the approach taken in Section
\ref{sec:Euclideanaction} by deriving the saddle point approximation
of transition amplitudes between axion conjugate momentum
eigenstates. Further, the imaginary part that, as we mentioned,
should be added to the on-shell action, Eq. \eqref{imagpart}, will
be shown to follow from a Fourier transformation relating axion
conjugate momentum eigenstates and axion field eigenstates. The
discussions and arguments presented in this Section are inspired by
\cite{Andres,Bergshoeff:2005zf}. We refer to
\cite{Lee:1988ge,Burgess:1989da,Brown:1989df} for related work in
four dimensions.

\subsection{Wick rotated path integrals and axions}

In classical field theory when going from the Lorentzian IIB
supergravity to Wick rotated Euclidean IIB supergravity we replace
$\chi'$ by $i\chi'$. Here, we will show that on the level of the
path integral $\chi'$ does not get replaced by $i\chi'$ when Wick
rotating the path integral.

Consider the path integral

\begin{equation}\label{pathintegral}
    \int\mathcal{D}T\,\mathcal{D}\mathcal{F}_9\,\mathcal{D}\chi'\,e^{iS\,[T,\mathcal{F}_9,\chi']}\,,
\end{equation}

\noindent with $S\,[T,\mathcal{F}_9,\chi']$ as given in
\eqref{firstorderchi'}. We do not include the metric in the
discussion concerning the path integral since the metric for the
instanton solutions is flat. The axion $\chi'$ in
\eqref{pathintegral} can be integrated over using the identity

\begin{equation}\label{deltafunctional}
    \int\mathcal{D}\chi'\,e^{-i\chi'd\mathcal{F}_9}=\delta[d\mathcal{F}_9]\,,
\end{equation}

\noindent where $\delta[d\mathcal{F}_9]$ is a delta-functional,
which implies that $d\mathcal{F}_9=0$.

The Wick rotated version of \eqref{pathintegral} is

\begin{equation}\label{euclideanpathintegral}
    \int\mathcal{D}T\,\mathcal{D}\mathcal{F}_9\,\mathcal{D}\chi'\,e^{-S_E\,[T,\mathcal{F}_9,\chi']}\,,
\end{equation}

\noindent where $S_E=-iS(\text{Wick rotated})$ is given by (leaving
out the metric)

\begin{align}
    S_E\,[T, \mathcal{F}_9, \chi']=&\int_{{\mathcal{M}}_{10}}\left(
    \frac{1}{2}\frac{1}{T^2-4\text{det}\,Q}
    \star dT\wedge dT+\right.\nonumber\\
    &\left.\frac{1}{2}\frac{1}{T^2-4\text{det}\,Q}\star \mathcal{F}_{9}\wedge\mathcal{F}_9
    +i\chi' d\mathcal{F}_{9}\right)\,.\label{euclideanfirstorderchi'}
\end{align}

\noindent In the path integral \eqref{euclideanpathintegral} we are
integrating over paths of field configurations with Dirichlet
boundary conditions for the fields $T$ and $\mathcal{F}_9$ while
free or no boundary conditions are imposed for the field $\chi'$.
These boundary conditions are the same as those imposed on the
variations of the action \eqref{firstorderchi'} with respect to $T$,
$\mathcal{F}_9$ and $\chi'$. The variation of $\chi'$ is entirely
free without any boundary conditions because it appears in
\eqref{euclideanfirstorderchi'} without a derivative.

Notice that $\chi'$ in \eqref{euclideanfirstorderchi'} has not been
replaced by $i\chi'$. Now in the Euclidean path integral $\chi'$ can
be again integrated out using the identity \eqref{deltafunctional}
which allows one to go to a second order formalism. If instead we
had replaced $\chi'$ by $i\chi'$ this would have no longer been
possible and the first order action in the Euclidean path integral
would not have been equivalent to an 8-form gauge theory anymore
since the Bianchi identity $d{\mathcal F}_9=0$ and its consequence
${\mathcal F}_9=d{\mathcal A}_8$ would not arise.

\subsection{The role of the moduli space}
Let us rewrite the last term in \eqref{euclideanfirstorderchi'} as
follows

\begin{equation}
    i\int_{{\mathcal{M}}_{10}}\chi'
    d\mathcal{F}_{9}=-i\int_{{\mathcal{M}_{10}}}d\chi'\wedge\mathcal{F}_9+
    i\int_{\partial{\mathcal{M}_{10}}}\chi'\mathcal{F}_9\,.
\end{equation}

\noindent If we require that the Euclidean path integral respects
the standard IIB symmetry $\chi'\rightarrow\chi'+b$ where $b$ is any
real number then we find that ${\mathcal F}_9$ should satisfy the
following boundary condition

\begin{equation}\label{quantization}
    b\int_{{\partial\mathcal{M}}_{10}}\mathcal{F}_9=2\pi
    n\qquad\text{with $n\in\mathbb{Z}$\,.}
\end{equation}

\noindent Since $b$ is arbitrary this means that
$\int_{{\partial\mathcal{M}}_{10}}\mathcal{F}_9$ has to vanish. This
would mean that there is no instanton present. If instead we only
require that the axion can undergo integer, in particular, unit
shifts $\chi'\rightarrow\chi'+1$ then we find that

\begin{equation}\label{quantization2}
    \int_{{\partial\mathcal{M}}_{10}}\mathcal{F}_9=2\pi
    n\qquad\text{with $n\in\mathbb{Z}$\,.}
\end{equation}

\noindent We conclude from this that instantons can only exist in
axion-dilaton theories whose moduli space is given by
\eqref{quantummodulispace}. We mention that the situation with the
7-brane solutions is in this respect entirely analogous. There the
arguments to use \eqref{quantummodulispace} are based on the
requirement of having 7-brane solutions with finite energy
\cite{Greene:1989ya}. The conclusion that one must factor the moduli
space $SL(2,\mathbb{R})/SO(2)$ by $SL(2,\mathbb{Z})$ in order to
even speak about instantons is clear from the path integral point of
view and does not follow from the classical field theory approach of
the previous two Sections.

\subsection{Integrating over $\mathcal{F}_9$}\label{integratingF9}

Instead of integrating out $\chi'$ we shall now integrate
\eqref{euclideanpathintegral} over $\mathcal{F}_9$. This is achieved
by defining a new 9-form $\mathcal{F}'_9$

\begin{equation}
    \mathcal{F}'_9=\mathcal{F}_9+i(T^2-4\text{det}\,Q)\star
    d\chi'\,.
\end{equation}

\noindent Such a shift of $\mathcal{F}_9$ in the imaginary direction
does not affect the integration in \eqref{euclideanpathintegral}.
The action \eqref{euclideanfirstorderchi'} now becomes

\begin{align}
    &S_E\,[T, \mathcal{F}'_9, \chi']=\int_{{\mathcal{M}}_{10}}\left(
    \frac{1}{2}\frac{1}{T^2-4\text{det}\,Q}
    \star dT\wedge dT+\frac{1}{2}\frac{1}{T^2-4\text{det}\,Q}\star \mathcal{F}'_{9}\wedge\mathcal{F}'_9
    \right.\nonumber\\
    &\left.+\frac{1}{2}(T^2-4\text{det}\,Q)\star d\chi'\wedge d\chi'\right)+
    i\int_{\partial{\mathcal{M}_{10}}}\chi'\mathcal{F}_9\,.
    \label{euclideanfirstorderchi'2}
\end{align}

\noindent Even though $\mathcal{F}_9$ appears in the boundary term
of \eqref{euclideanfirstorderchi'2} the $\mathcal{F}'_9$ integral is
a Gaussian as we are integrating over $\mathcal{F}'_9$ with
Dirichlet boundary conditions. The $\mathcal{F}_9$ in the boundary
term is not integrated over, but is fixed by the identification
$\chi'\sim\chi'+1$, see Eq. \eqref{quantization2}. Integrating over
$\mathcal{F}'_9$ we find the following path integral

\begin{equation}\label{euclideanpathintegral2}
    \int_F(T^2-4\text{det}\,Q)^{1/2}\mathcal{D}T\,\mathcal{D}\chi'\,e^{-\tilde S_E\,[T,\chi']}\,,
\end{equation}

\noindent where $F$ below the integral sign means to indicate that
we are only integrating over the paths of field configurations that
are within the fundamental domain of the quantum moduli space
\eqref{quantummodulispace}. From now on this will always be assumed
and the label $F$ will be suppressed. The integration
measure\footnote{In terms of $\tau$ and $\bar\tau$ the integration
measure would be
$(\text{Im}\,\tau)^{-2}\mathcal{D}\tau\mathcal{D}\bar\tau$, which is
$PSL(2,\mathbb{Z})$ invariant.} now contains the factor
$(T^2-4\text{det}\,Q)^{1/2}$ and the Euclidean action $\tilde
S_E\,[T,\chi']$ is given by

\begin{equation}
    \tilde S_E\,[T,\chi']=\int_{{\mathcal{M}}_{10}}\left(
    \frac{1}{2}\frac{1}{T^2-4\text{det}\,Q}
    \star dT\wedge dT+\frac{1}{2}(T^2-4\text{det}\,Q)\star d\chi'\wedge d\chi'\right)+
    i\int_{\partial{\mathcal{M}_{10}}}\chi'\mathcal{F}_9\,.
    \label{euclideanfirstorderchi'3}
\end{equation}

\subsection{Splitting the $\chi'$ integration into bulk and boundary integrations}

We split up the integration over $\chi'$ into two pieces: the
integration over bulk $\chi'$ field configurations and the
integration over boundary $\chi'_{\partial}$ field configurations.
The bulk $\chi'$ field configurations will be denoted by the same
symbol as was used in the previous Subsections. Since we now
explicitly write $\chi'_{\partial}$ for the boundary values this
should cause no confusion. This split is most easily done using
Dirichlet boundary conditions for the paths appearing in the path
integral over the bulk $\chi'$ field configurations. If we do this
then we can write for \eqref{euclideanpathintegral2}

\begin{equation}\label{euclideanpathintegral3}
    \int(T^2-4\text{det}\,Q)^{1/2}\mathcal{D}T\,\mathcal{D}\chi'\mathcal{D}\chi'_{\partial}
    \,e^{-S_E\,[T,\chi',\chi'_{\partial}]}\,,
\end{equation}

\noindent with Dirichlet boundary conditions on the integrations
over $T$ and $\chi'$. The action appearing in
\eqref{euclideanpathintegral3} is given by

\begin{align}
    S_E\,[T,\chi',\chi'_{\partial}]=&\int_{{\mathcal{M}}_{10}}\left(
    \frac{1}{2}\frac{1}{T^2-4\text{det}\,Q}
    \star dT\wedge dT+\frac{1}{2}(T^2-4\text{det}\,Q)\star d\chi'\wedge d\chi'\right)\nonumber\\
    &+i\int_{\partial\mathcal{M}_{10}}\chi'_{\partial}\,\mathcal{F}_9\,.
    \label{euclideanfirstorderchi'4}
\end{align}

\noindent The variation of the bulk part of
\eqref{euclideanfirstorderchi'4} with respect to $T$ and $\chi'$
satisfying Dirichlet boundary conditions produces the standard
(non-Wick rotated) IIB axion-dilaton equations of motion.

\subsection{Tunneling interpretation}

In this Subsection we will discuss what is precisely computed by the
Euclidean path integral \eqref{euclideanpathintegral}, \emph{i.e.}
by \eqref{euclideanpathintegral3}.

We would like to interpret \eqref{euclideanpathintegral3} in terms
of matrix elements describing a tunneling process from an initial
($t=-\infty$) time-like hypersurface $\Sigma_i$ to a final
($t=+\infty$) time-like hypersurface $\Sigma_f$. The time-like
hypersurfaces $\Sigma_i$ and $\Sigma_f$ constitute surfaces on which
field operator states exist. In order to describe this within the
space-time $\mathcal{M}_{9,1}$ we add to it spatial infinity as a
point. Hence we consider $\mathcal{M}_{9,1}\cup\{r=\infty\}$, where
$r$ is a radial coordinate. The topology of this one-point
compactified space-time is given by $\mathbb{R}\times S^9$ whose
boundary $\partial\left(\mathcal{M}_{9,1}\cup\{r=\infty\}\right)$ is
given by the disjoint union $\Sigma_i\cup\Sigma_f$ where the initial
and final time-like hypersurfaces have the topology of $S^9$.

The instanton charge $\int_{\partial\mathcal{M}_{10}}\mathcal{F}_9$
that appears in the imaginary part of Eq.
\eqref{euclideanfirstorderchi'4} is equal to
$\int_{\Sigma_f}\mathcal{F}_{9}^f-\int_{\Sigma_i}\mathcal{F}^i_9$.
Multiplying this equality by $\chi'_{\partial}$ we can write

\begin{equation}\label{instantonchargeIF}
i\int_{\partial\mathcal{M}_{10}}\chi'_{\partial}\,\mathcal{F}_9=
i\int_{\Sigma_f}\chi'_{\partial}\,\mathcal{F}^f_9-i\int_{\Sigma_i}\chi'_{\partial}\,\mathcal{F}^i_9
\,,
\end{equation}

\noindent where the values of the axion $\chi'$ on the initial and
final timelike hypersurfaces $\Sigma_i$ and $\Sigma_f$ are the same:
$\chi'_i=\chi'_f=\chi'_{\partial}$. In the following we will write
$\chi'_{\partial}=\chi'_{\infty}$. Further we have

\begin{equation}\label{rewritingimagipart}
    \int\mathcal{D}\chi'_{\partial}e^{-i\int_{\partial\mathcal{M}_{10}}\chi'_{\partial}\,\mathcal{F}_9}=
\int\mathcal{D}\chi'_{i}\mathcal{D}\chi'_{f}\delta(\chi'_i-\chi'_f)
e^{-i\int_{\Sigma_f}\chi'_{f}\,\mathcal{F}^f_9+i\int_{\Sigma_i}\chi'_{i}\,\mathcal{F}^i_9}\,.
\end{equation}
The boundary states in \eqref{euclideanpathintegral3} at
$\Sigma_{i,f}$ satisfy \eqref{quantization2} and
\eqref{instantonchargeIF} and so the $\mathcal{F}_9^{i,f}$ boundary
data are on-shell.

We will now show that one can use the duality relation
\eqref{SU(1,1)axion} restricted to the surfaces $\Sigma_{i,f}$ to
interpret the boundary data $\mathcal{F}_9^{i,f}$ of Eq.
(\ref{rewritingimagipart}) in terms of the axion momentum, or
equivalently, in terms of the Noether charge density associated with
the axion shift symmetry. Note that the time component of the
Noether current (charge density) is equal to the axion $\chi'$
canonical momentum $\pi'$ obtained by varying the Lagrangian
(\ref{pqrIIBaction}) with respect to $\partial_0\chi'$
\begin{equation}\label{moment}
\pi'=\frac{\delta\mathcal{L}}{\delta(\partial_0\chi')}=(T^2-4\text{det}\,Q)\,\partial_0\chi'=J_N^0\,.
\end{equation}

Let us consider tunneling between canonical momentum eigenstates of
the axion $\chi'$ (or equivalently between its Noether charge
eigenstates) from the initial surface $\Sigma_i$ to the final
surface $\Sigma_f$. These are described by the following matrix
element

\begin{equation}\label{mepi}
    \lim_{\Delta T\to\infty}\langle \pi'_f\mid e^{-H\Delta T}\mid \pi'_i\rangle\,,
\end{equation}

\noindent where $\Delta T$ is the Wick rotated time interval between
$\Sigma_i$ and $\Sigma_f$, $H$ is the axion--dilaton Hamiltonian
which can be obtained from the (flat metric) action
(\ref{pqrIIBaction}) by the Legendre transformation and $\pi'_{i,f}$
are the initial and final momenta of the axion.

The matrix element (\ref{mepi}) is related by a Fourier
transformation to the matrix element describing the transition
between two boundary eigenstates $\chi'_i$ and $\chi'_f$ of the
axion. Namely, (for $\Delta T\rightarrow\infty$) we have
\begin{equation}\label{tunneling2}
    \langle \pi'_f\mid e^{-H\Delta T}\mid \pi'_i\rangle=
\int\mathcal{D}\chi'_{i}\mathcal{D}\chi'_{f}
e^{-i\int_{\Sigma_f}\chi'_{f}\,\pi'_f+i\int_{\Sigma_i}\chi'_{i}\,\pi'_i}
\langle \chi'_f\mid e^{-H\Delta T}\mid \chi'_i\rangle\,,
\end{equation}

\noindent where

\begin{equation}\label{axionmatrixelementIF}
    \langle \chi'_f\mid e^{-H\Delta T}\mid \chi'_i\rangle=\langle \chi'_f\mid e^{-H\Delta T}\mid \chi'_f\rangle
    \delta(\chi'_i-\chi'_f)\,.
\end{equation}

\noindent We see that no tunneling takes place between vacua for
which $\chi'_i\neq\chi'_f$. This means that the value of $\chi'$ at,
say, $t=+\infty$ acts as a superselection parameter, like the theta
parameter in Yang--Mills theory. Hence, physical processes in vacua
with different values of $\chi'_f$ are not correlated.

The matrix element appearing on the right-hand side of Eq.
\eqref{axionmatrixelementIF} is given by

\begin{equation}\label{matrixelement}
    \langle \chi'_f\mid e^{-H\Delta T}\mid \chi'_{f}\rangle=
    \int(T^2-4\text{det}\,Q)^{1/2}\mathcal{D}T\,\mathcal{D}\chi'\,
    e^{-S_E\,[T,\chi']}\,,
\end{equation}

\noindent with Dirichlet boundary conditions on the integrations
over $T$ and $\chi'$ and

\begin{equation}\label{ada}
S_E\,[T,\chi']=\int_{{\mathcal{M}}_{10}}\left(
    \frac{1}{2}\frac{1}{T^2-4\text{det}\,Q}
    \star dT\wedge dT+\frac{1}{2}(T^2-4\text{det}\,Q)\star d\chi'\wedge d\chi'\right)\,.
\end{equation}
We now compare Eqs. (\ref{tunneling2})--(\ref{ada}) with
(\ref{euclideanpathintegral3})--(\ref{rewritingimagipart}). Eqs.
(\ref{tunneling2})--(\ref{ada}) taken together provide a closed
expression for the matrix element on the left hand-side of Eq.
\eqref{tunneling2}. On the other hand Eqs.
(\ref{euclideanpathintegral3})--(\ref{rewritingimagipart}) provide
an expression for the path integral in
\eqref{euclideanpathintegral3}. In order that
\eqref{euclideanpathintegral3} computes a physical quantity, namely
the matrix element of \eqref{tunneling2}, we choose the boundary
values of ${\mathcal F}_9$ to be associated with the boundary values
of the $\chi'$ canonical momentum (\ref{moment}) via the duality
relation (\ref{SU(1,1)axion})
\begin{equation}\label{Noethercharge}
    \int_{\Sigma_{i,f}}\mathcal{F}_9^{i,f}=\int_{\Sigma_{i,f}}\star
    (T^2-4\text{det}\,Q)d\chi'=\int_{\Sigma_{i,f}}J_N^0\,d\Omega_9=
    \int_{\Sigma_{i,f}}\pi'_{i,f}\,d\Omega_9\,,
\end{equation}

\noindent where $d\Omega_9$ denotes the integration measure of the
unit 9-sphere.

Using the inverse Fourier transform we have

\begin{equation}\label{axionmatrixelement}
\langle \chi'_f\mid e^{-H\Delta T}\mid \chi'_{i}\rangle=\int\mathcal{D}\pi'_{i}\mathcal{D}\pi'_{f}
e^{i\int_{\Sigma_f}\chi'_{f}\,\pi'_f-i\int_{\Sigma_i}\chi'_{i}\,\pi'_i}
\langle \pi'_f\mid e^{-H\Delta T}\mid \pi'_i\rangle\,,
\end{equation}

\noindent where now

\begin{equation}\label{conjmommatrixelem}
\langle \pi'_f\mid e^{-H\Delta T}\mid \pi'_i\rangle=
\int\mathcal{D}T\,\mathcal{D}\mathcal{F}_9\,\delta[d\mathcal{F}_9]\,e^{-S_E\,[T,\mathcal{F}_9]}\,,
\end{equation}

\noindent with Dirichlet boundary conditions imposed on the
integrations over $T$ and $\mathcal{F}_9$ and where

\begin{equation}\label{euclideanaction9form}
    S_E\,[T, \mathcal{F}_9]=\int_{{\mathcal{M}}_{10}}\left(
    \frac{1}{2}\frac{1}{T^2-4\text{det}\,Q}
    \star dT\wedge dT+\frac{1}{2}\frac{1}{T^2-4\text{det}\,Q}\star \mathcal{F}_{9}\wedge\mathcal{F}_9
    \right)\,.
\end{equation}

\subsection{Saddle point approximation}

The saddle point approximation of $\langle \chi'_f\mid e^{-H\Delta
T}\mid \chi'_{f}\rangle$ can be obtained using Eqs.
\eqref{axionmatrixelement}, \eqref{conjmommatrixelem} and
\eqref{euclideanaction9form}. For a single instanton of charge $n$
we have

\begin{equation}\label{saddleapprox}
    \langle \chi'_f\mid e^{-H\Delta
T}\mid \chi'_{f}\rangle\simeq
N e^{i\int_{\Sigma_f}\chi'_{f}\,\pi'_f-i\int_{\Sigma_i}\chi'_{f}\,\pi'_i}e^{-S_E\,[T, \mathcal{F}_9]}\mid_{\text{on-shell}}\,,
\end{equation}

\noindent where $N$ is a prefactor that we will not attempt to
evaluate. On the mass shell we have

\begin{equation}\label{instantoncharge}
    \int_{\Sigma_f}\chi'_{f}\,\pi'_fd\Omega_9-\int_{\Sigma_i}\chi'_{f}\,\pi'_id\Omega_9=2\pi n\chi'_\infty\,,
\end{equation}

\noindent which follows from Eqs. \eqref{quantization2} and
\eqref{instantonchargeIF}. Further, on-shell and outside the
Q-instanton source $S_E\,[T, \mathcal{F}_9]=S_E\,[T,
d\mathcal{A}_8]$. The on-shell action can be written as the sum of a
quadratic term and a rest term as

\begin{equation}\label{actionassumsquares}
S_{E}\,[T, d\mathcal{A}_8]=\frac{1}{2}\int_{{\mathcal{M}}_{10}}\frac{1}{T^2-4\text{det}\,Q}\star\left(dT
\mp \star \mathcal{F}_9) \wedge \, (dT
\mp \star \mathcal{F}_9 \right)\pm G\,,
\end{equation}

\noindent with $G$ given by

\begin{equation}
G=\int_{{\mathcal{M}}_{10}}\frac{1}{T^2-4\text{det}\,Q} dT \wedge
\,\mathcal{F}_9=
-\int_{\partial{\mathcal{M}}_{10}}\frac{1}{4\sqrt{\text{det}\,Q}}\log
\left(\frac{T+2\sqrt{\text{det}\,Q}}{T-2\sqrt{\text{det}\,Q}}\right)\mathcal{F}_9
\,,
\end{equation}

\noindent where
$\partial{\mathcal{M}}_{10}=\partial{\mathcal{M}}_{\infty}+\partial{\mathcal{M}}_{0}$.
The boundaries $\partial{\mathcal{M}}_{\infty}$ and
$\partial{\mathcal{M}}_{0}$ are, respectively, the 9-sphere at
infinity and around the origin where the field strength $F_9$ fails
to be exact (the location of its magnetic source). However because
at $\vert\vec x\vert=0$ the field $T$ blows up, the value of $G$ is
zero at this point and only the boundary at infinity contributes.
The first term in the action \eqref{actionassumsquares} is positive
definite. We thus have the following Bogomol'nyi bound for field
configurations respecting the symmetries of the Q(-1)-brane solution

\begin{equation}
    S_I\ge\pm G\,.
\end{equation}

Solutions that satisfy the Bogomol'nyi bound must have the property
that

\begin{equation}\label{Bogomolnyibound}
dT = \pm \star \mathcal{F}_9\,.
\end{equation}

\noindent For such configurations the on-shell value of the action
is given by

\begin{equation}\label{action boundary term}
S_E\,[T,d\mathcal{A}_8]\mid_{\text{on-shell}}=-G=
\frac{\pi \vert n\vert}{2\sqrt{\text{det}\,Q}}\log\left(\frac{T_\infty+2\sqrt{\text{det}\,Q}}{T_\infty-2\sqrt{\text{det}\,Q}}
\right)\,,
\end{equation}

\noindent where $T_\infty>2\sqrt{\text{det}\,Q}$ is the asymptotic
value of $T$. The result \eqref{action boundary term} agrees with
\eqref{onshellactionQ} and provides a saddle point approximation of
the matrix element of a transition between axion charge eigenstates
(or conjugate momentum eigenstates).

Using Eqs. \eqref{instantoncharge} and \eqref{action boundary term}
we find for the saddle point approximation \eqref{saddleapprox},

\begin{equation}\label{saddleapproxwithimagpart}
\langle \chi'_f\mid e^{-H\Delta
T}\mid \chi'_{f}\rangle\simeq
Ne^{2\pi ni\chi'_\infty-2\pi \vert n\vert\text{Im}\mathcal{T}_{\infty}}=\left\{\begin{array}{ll}
Ne^{2\pi ni\mathcal{T}_{\infty}}& \quad\text{for $n>0$}\,,\\
Ne^{2\pi ni\bar{\mathcal{T}}_{\infty}}& \quad\text{for $n<0$}\,.
\end{array}\right.
\end{equation}

\noindent The case $n>0$ corresponds to the Q-instanton whereas
$n<0$ corresponds to the anti-Q-instanton. We thus see that adding
the term \eqref{imagpart} to the action \eqref{onshellactionQ} leads
to a saddle point approximation of the matrix element of the
transition between axion eigenstates
$\chi'_i=\chi'_f=\chi'_{\infty}$. The result
\eqref{saddleapproxwithimagpart} will be used in the next Section to
argue that the $\mathcal{R}^4$ terms near the points $i$ and $\rho$
of figure \ref{fundamentaldomain} receive contributions from
Q-instantons.

\section{Q-instanton contributions to the $\mathcal{R}^4$ terms}\label{sec:Rto4}

The $\mathcal{R}^4$ terms are those terms in the effective action
that are of order $(\alpha')^3$ relative to the Einstein--Hilbert
term. In \cite{Green:1997tv} it is argued that the part of the
$\mathcal{R}^4$ terms that only contains derivatives of the metric
is multiplied by a $PSL(2,\mathbb{Z})$ invariant real-analytic
modular form, a generalized Eisenstein series. Such functions are
eigenfunctions of the Laplace operator on the hyperbolic plane. In
\cite{Green:1998by} it is shown that this picture is confirmed by
requiring supersymmetry at the order $(\alpha')^3$ relative to the
Einstein--Hilbert term. The $\mathcal{R}^4$ terms contain besides
derivatives of the metric also contributions involving terms with
derivatives of the other bosonic fields of the type IIB theory. For
the NSNS fields and the RR 0-form a conjectured $SL(2,\mathbb{Z})$
invariant $\mathcal{R}^4$ term is proposed in
\cite{Kehagias:1997cq}. Here we will only consider the part of the
$\mathcal{R}^4$ terms that involves derivatives of the metric and
that can be obtained by considering on-shell amplitudes for four
graviton scattering. We write \cite{Kehagias:1997cq}

\begin{equation}\label{Rtofour}
    \mathcal{R}^4=f(\tau,\bar\tau)\,\left(t_8^{abcdefgh}t_8^{mnpqrstu}+\frac{1}{8}
    \epsilon_{10}^{abcdefghij}\epsilon_{10\quad\quad\;\;\; ij}^{mnpqrstu}\right)R_{abmn}R_{cdpq}
    R_{efrs}R_{ghtu}+\cdots\,,
\end{equation}

\noindent where $t_8$ is defined in \cite{Schwarz:1982jn},
$\epsilon_{10}$ is the 10-dimensional Levi-Civit\`{a} tensor and
$R_{abmn}$ is the Riemann tensor. The dots indicate that there are
more contributions to $\mathcal{R}^4$.

The function $f(\tau,\bar\tau)$, a generalized Eisenstein series,
has the form

\begin{equation}\label{functionf}
    f(\tau,\bar\tau)=\sum_{(p,n)\neq (0,0)}\frac{\tau_2^{3/2}}{\vert p+n\tau\vert^3}\,,
\end{equation}

\noindent where $\tau=\tau_1+i\tau_2$ and the sum is over all
integers $p,n\in\mathbb{Z}$ except when both $p$ and $n$ are zero.
In order to see the contributions coming from single
multiply--charged D- and anti-D-instantons one writes $f$ as a
Fourier series in $\tau_1=\chi$. We have \cite{Terras:1985}

\begin{align}
    f(\tau,\bar\tau)&=2\zeta(3)\tau_2^{3/2}+\frac{2\pi^2}{3}\tau_2^{-1/2}+8\pi\tau_2^{1/2}\sum_{m\neq 0}\sum_{n=1}^{\infty}\vert\frac{m}{n}\vert e^{2\pi i mn\tau_1}\,K_1(2\pi\vert mn\vert\tau_2)\nonumber\\
    & =2\zeta(3)\tau_2^{3/2}+\frac{2\pi^2}{3}\tau_2^{-1/2}+8\pi\tau_2^{1/2}\sum_{k=1}^{\infty}k\,\sigma_{-2}(k)\left(
    e^{2\pi i k\tau_1}+e^{-2\pi i k\tau_1}\right)\,K_1(2\pi k\tau_2)\,,\label{Fseries}
\end{align}

\noindent with $\sigma_{-2}(k)$ given by

\begin{equation}
    \sigma_{-2}(k)=\sum_{d\vert k}d^{-2}\,,
\end{equation}

\noindent where the sum is over all positive divisors $d$ of $k$.
The expression \eqref{Fseries} is a cosine series with coefficients
$16\pi\tau_2^{1/2}\, k\,\sigma_{-2}(k)\,K_1(2\pi k\tau_2)$, where
$K_1$ is the modified Bessel function of the second kind. The
$\tau_1$ independent terms in \eqref{Fseries} do not come from
D-instantons, instead they come from an $(\alpha')^3$ tree level and
a one-loop effect in the four graviton amplitude
\cite{Green:1997tv}.

In order to see the contribution from single multiply--charged D-
and anti-D-instantons one considers \eqref{Fseries} close to
$\tau_0=i\infty$, \emph{i.e.} in the limit
$\tau_2\rightarrow\infty$. Using that for $x\rightarrow\infty$ we
have $K_1(x)=\sqrt{\tfrac{\pi}{2x}}e^{-x}(1+\ldots)$ we find that at
the leading order in the limit $\tau_2\rightarrow\infty$

\begin{equation}\label{asymptFseries}
f(\tau,\bar\tau)\approx 2\zeta(3)\tau_2^{3/2}+\frac{2\pi^2}{3}\tau_2^{-1/2}+4\pi\sum_{k=1}^{\infty}k^{1/2}\,\sigma_{-2}(k)\left(
    e^{2\pi i k\tau}+e^{-2\pi i k\bar\tau}\right)\,.
\end{equation}
In the exponents of (\ref{asymptFseries}) one recognizes the
D-instanton action (\ref{onshelld1}).

The Q-instantons of this paper preserve the same supersymmetries as
the D-instanton. It is therefore expected that they will also
contribute to the function $f(\tau,\bar\tau)$. To justify this
argument, in the remainder of this Section we shall Fourier expand
the function $f$ in terms of $\chi'$ and compute the Fourier
coefficients which will be functions of $T$. This will result in an
exact expression for $f$ which is analogous to Eq. \eqref{Fseries}.
Schematically we write

\begin{equation}
    f(T,\chi')=\sum_{n=-\infty}^{\infty}c_{n}(T)e^{2\pi in\chi'}\,,
\end{equation}

\noindent where $c_n(T)$ are the Fourier coefficients. This series
is manifestly invariant under $\chi'\rightarrow\chi'+1$ and provides
us with the behavior of $f$ near $\tau=\tau_0$. It will be shown
that $f$ consists of a $\chi'$ independent part and of a cosine
series that corresponds to an infinite sum of single
multiply--charged Q- and anti-Q-instantons.

We shall expand the function $f$ around the fixed points
$\tau_0=i,\rho$ of the axion--dilaton moduli space. To this end it
will prove convenient to introduce what we refer to as the
$(\eta,\varphi)$ coordinate system which is defined by the relation

\begin{equation}\label{etavarphisystem}
    \frac{\tau-\tau_0}{\tau-\bar\tau_0}=e^{i\varphi}\tanh\frac{\eta}{2}\,,
\end{equation}

\noindent where $\varphi$ and $\eta$ are related to $T$ and $\chi'$
(see Eqs. \eqref{tautoT} and \eqref{mathcalT}) as
follows\footnote{The transformation from $T$ to $\eta$ is such that
the kinetic term for $T$ in the action becomes canonically
normalized, \emph{i.e.} we have

\begin{equation}
    \frac{\star dT\wedge dT}{T^2-4\text{det}\,Q}=\star d\eta\wedge d\eta\,.
\end{equation}}

\begin{align}
    \varphi &=2\sqrt{\text{det}\,Q}\chi'\qquad\text{where}\quad 0\le\varphi<2\pi\,, \label{defvarphi}\\
    \tanh\frac{\eta}{2} &=e^{-2\sqrt{\text{det}\,Q}\,\text{Im}\,\mathcal{T}}\,\quad\text{where}\quad 0<\eta<\infty\,.\label{defeta}
\end{align}

\noindent Substituting \eqref{etavarphisystem} into
\eqref{functionf} we obtain

\begin{equation}\label{fasfunctionofetavarphi}
    f(\eta,\varphi)=\sum_{(p,n)\neq(0,0)}\frac{(\text{Im}\,\tau_0)^{3/2}}{\vert p+n\tau_0\vert^3}
    \frac{1}{\left(\cosh\eta+\sinh\eta\cos(\varphi+\beta(p,n;\tau_0))\right)^{3/2}}\,,
\end{equation}

\noindent where $\beta(p,n;\tau_0)$ is defined by

\begin{align}
    \cos\beta(p,n;\tau_0)&=\frac{n^2(\text{Im}\,\tau_0)^2-(p+n\text{Re}\,\tau_0)^2}{n^2(\text{Im}\,\tau_0)^2+(p+n\text{Re}\,\tau_0)^2}\,,\label{cosbeta}\\
    \sin\beta(p,n;\tau_0)&=\frac{2n\,\text{Im}\,\tau_0\,(p+n\text{Re}\,\tau_0)}{n^2(\text{Im}\,\tau_0)^2+(p+n\text{Re}\,\tau_0)^2}\,.\label{sinbeta}
\end{align}

From the definition of $\varphi$ in terms of $\chi'$ it follows that
the invariance of $f(T,\chi')$ under $\chi'\rightarrow\chi'+1$
implies the invariance of $f(\eta,\varphi)$ under
$\varphi\rightarrow\varphi+2\sqrt{\text{det}\,Q}$. Hence, we make
the following Fourier series decomposition of $f(\eta,\varphi)$

\begin{equation}\label{Fourierdecomposition}
    f(\eta,\varphi)=\sum_{m=-\infty}^{\infty}a_{\tfrac{\pi m}{\sqrt{\text{det}\,Q}}}(\eta)e^{\tfrac{\pi}{\sqrt{\text{det}\,Q}}mi\varphi}\,.
\end{equation}

\noindent The Fourier coefficients $a_{\tfrac{\pi
m}{\sqrt{\text{det}\,Q}}}$ are given by

\begin{equation}\label{Fouriercoefficient}
    a_{\tfrac{\pi m}{\sqrt{\text{det}\,Q}}}(\eta)=\frac{1}{2\sqrt{\text{det}\,Q}}\int_0^{2\sqrt{\text{det}\,Q}}d\varphi\,f(\eta,\varphi)e^{-\tfrac{\pi}{\sqrt{\text{det}\,Q}}mi\varphi}\,.
\end{equation}

\noindent By using \eqref{fasfunctionofetavarphi} and by shifting
the integration over $\varphi$ in \eqref{Fouriercoefficient} to an
integration over $\theta=\varphi+\beta(p,n;\tau_0)$ the Fourier
coefficients $a_{\tfrac{\pi m}{\sqrt{\text{det}\,Q}}}$ can be
written as

\begin{align}\label{masterformulaFouriercoefficients}
    a_{\tfrac{\pi m}{\sqrt{\text{det}\,Q}}}(\eta)=&\frac{1}{2\sqrt{\text{det}\,Q}}\sum_{(p,n)\neq(0,0)}\frac{(\text{Im}\,\tau_0)^{3/2}}{\vert p+n\tau_0\vert^3}\,e^{\tfrac{\pi}{\sqrt{\text{det}\,Q}}mi\beta(p,n;\tau_0)}\times\nonumber\\
& \times
\int_{\beta(p,n;\tau_0)}^{2\sqrt{\text{det}\,Q}+\beta(p,n;\tau_0)}d\theta\frac{e^{-\tfrac{\pi}{\sqrt{\text{det}\,Q}}mi\theta}}{\left(\cosh\eta+\sinh\eta\cos\theta\right)^{3/2}}\,.
\end{align}

We will further evaluate \eqref{masterformulaFouriercoefficients}
for the cases $\tau_0=i$ and $\tau_0=\rho$ separately. We start with
the case $\tau_0=i$. In Table \ref{tableorbifolddata} we presented
some data regarding the orbifold points $\tau_0=i,\rho$. We found
that for $\tau_0=i$ we have $2\sqrt{\text{det}\,Q}=\pi$. From Eqs.
\eqref{cosbeta} and \eqref{sinbeta} specified to the case $\tau_0=i$
we derive the following two identities

\begin{align}
    & \beta(-n,p;i)=\pi+\beta(p,n;i)\,,\label{shiftwithpi}\\
    & \beta(n,p;i)=\pi-\beta(p,n;i)\,.\label{interchangingpnfori}
\end{align}

\noindent Using the identity \eqref{shiftwithpi} we can write

\begin{align}
    &\sum_{(p,n)\neq(0,0)}\frac{e^{2mi\beta(p,n;i)}}{(p^2+n^2)^{3/2}}\,\int_{\beta(p,n;i)}^{\pi+\beta(p,n;i)}d\theta\frac{e^{-2mi\theta}}{\left(\cosh\eta+\sinh\eta\cos\theta\right)^{3/2}}=\nonumber\\
    &\sum_{(p,n)\neq(0,0)}\frac{e^{2mi\beta(p,n;i)}}{(p^2+n^2)^{3/2}}\,\int_{\pi+\beta(p,n;i)}^{2\pi+\beta(p,n;i)}d\theta\frac{e^{-2mi\theta}}{\left(\cosh\eta+\sinh\eta\cos\theta\right)^{3/2}}=\\
    &\frac{1}{2}\sum_{(p,n)\neq(0,0)}\frac{e^{2mi\beta(p,n;i)}}{(p^2+n^2)^{3/2}}\,\int_{0}^{2\pi}d\theta\frac{e^{-2mi\theta}}{\left(\cosh\eta+\sinh\eta\cos\theta\right)^{3/2}}\,,\nonumber
\end{align}

\noindent where in the last equality we took the average of the
first two lines and used the property
$\int_{\beta}^{2\pi+\beta}=\int_{\beta}^0+\int_{0}^{2\pi}+\int_{2\pi}^{2\pi+\beta}=\int_0^{2\pi}$
because the integrand is periodic with the period $2\pi$. We thus
find that the Fourier coefficients
\eqref{masterformulaFouriercoefficients} take the form

\begin{equation}\label{Fouriercoefficientsfori}
    a_{2m}(\eta)=\frac{1}{2\pi}\sum_{(p,n)\neq(0,0)}\frac{e^{2mi\beta(p,n;i)}}{(p^2+n^2)^{3/2}}\,\int_{0}^{2\pi}d\theta\frac{\cos(2m\theta)}{\left(\cosh\eta+\sinh\eta\cos\theta\right)^{3/2}}\,,
\end{equation}

\noindent where the integral from $0$ to $2\pi$ that involves
$\sin(2m\theta)$ vanished.

The identity \eqref{interchangingpnfori} can be used to show that
the sum preceding the integral in \eqref{Fouriercoefficientsfori}
satisfies

\begin{equation}
    \sum_{(p,n)\neq(0,0)}\frac{e^{2mi\beta(p,n;i)}}{(p^2+n^2)^{3/2}}=\sum_{(p,n)\neq(0,0)}\frac{e^{-2mi\beta(p,n;i)}}{(p^2+n^2)^{3/2}}\,,
\end{equation}

\noindent so that $a_{2m}(\eta)=a_{-2m}(\eta)$. The latter property
implies that the Fourier expansion \eqref{Fourierdecomposition}
becomes the following cosine series

\begin{equation}\label{cosineseriesf}
    f(\eta,\varphi)=a_0(\eta)+\sum_{m=1}^{\infty}a_{2m}(\eta)\left(e^{2mi\varphi}+e^{-2mi\varphi}\right)\,.
\end{equation}

The integral in \eqref{Fouriercoefficientsfori} is the integral
representation (up to a factor) of a toroidal function, denoted by
$P^{2m}_{1/2}(\cosh\eta)$. Toroidal or ring functions are special
cases of the associated Legendre functions. We have \cite{Bateman}

\begin{equation}\label{integralrepresentation}
\int_{0}^{2\pi}d\theta\frac{\cos(n\theta)}{\left(\cosh\eta+\sinh\eta\cos\theta\right)^{3/2}}=2\pi (-1)^n\frac{\Gamma(\tfrac{3}{2}-n)}{\Gamma(\tfrac{3}{2})}\,P^n_{1/2}(\cosh\eta)\,.
\end{equation}

\noindent The functions $P^{n}_{1/2}(\cosh\eta)$ for
$n=0,1,2,\ldots$ can be written in terms of a hypergeometric
function as follows \cite{Bateman}

\begin{equation}\label{functionalrelationF}
    P_{1/2}^n(\cosh\eta)=\frac{1}{2^n}\frac{\Gamma(\tfrac{3}{2}+n)}{\Gamma(n+1)\Gamma(\tfrac{3}{2}-n)}\,\sinh^n\eta\,F(\tfrac{n}{2}+\tfrac{3}{4},\tfrac{n}{2}-\tfrac{1}{4};n+1;-\sinh^2\eta)\,.
\end{equation}

Substituting Eqs. \eqref{integralrepresentation} and
\eqref{functionalrelationF} into \eqref{Fouriercoefficientsfori} we
see that the function $f(\eta,\varphi)$, Eq. \eqref{cosineseriesf},
around the point $\tau_0=i$ can be written as the following Fourier
series

\begin{align}
    &f(\eta,\varphi)=\sum_{(p,n)\neq(0,0)}\frac{1}{(p^2+n^2)^{3/2}}\,F(\tfrac{3}{4},-\tfrac{1}{4};1;-\sinh^2\eta)+\sum_{m=1}^{\infty}\sum_{(p,n)\neq(0,0)}\frac{e^{2mi\beta(p,n;i)}}{(p^2+n^2)^{3/2}}\,\times\nonumber\\
    &\times\frac{1}{2^{2m}}\frac{\Gamma(\tfrac{3}{2}+2m)}{\Gamma(2m+1)\Gamma(\tfrac{3}{2})}\,\sinh^{2m}\eta\,F(m+\tfrac{3}{4},m-\tfrac{1}{4};2m+1;-\sinh^2\eta)\left(e^{2mi\varphi}+e^{-2mi\varphi}\right)\,.\label{finalresultfori}
\end{align}

\noindent From Eqs. \eqref{defvarphi} and \eqref{defeta} we know
that

\begin{equation}
    \varphi=\pi\chi' \qquad\text{and}\qquad
    \sinh^2\eta=\frac{T^2-4\text{det}\,Q}{4\text{det}\,Q}\qquad\text{with}\qquad
    \sqrt{\text{det}\,Q}=\frac{\pi}{2}\,.
\end{equation}

\noindent The Fourier series \eqref{finalresultfori} in terms of
$\chi'$ and $T^2-4\text{det}\,Q$ associated with the fixed point
$\tau_0=i$ is analogous to the Fourier series expansion
\eqref{Fseries} around the point $\tau_0=i\infty$ in terms of
$\tau_1=\chi$ and $\tau_2=\text{Im}\,\tau=e^{-\phi}$.

In order to make manifest the Q- and anti-Q-instanton contributions
to the function $f$ we consider the expansion
\eqref{finalresultfori} at the leading order around the point
$\eta=0$ (that corresponds to a singular point of the associated
Legendre function $P^{2m}_{1/2}(\cosh\eta)$). Note that, by virtue
of the relation \eqref{etavarphisystem}, the point $\eta=0$
corresponds to $\tau=i$. Using that at leading order

\begin{equation}
    \sinh^2\eta\approx 4\,e^{-2\pi\text{Im}\,\mathcal{T}}\,,
\end{equation}

\noindent we find that at this order\footnote{This can alternatively
be derived by using that $P_{1/2}^{n}(\cosh\eta)$ can also be
written as
\begin{equation}
P_{1/2}^{n}(\cosh\eta)=\frac{\Gamma(\tfrac{3}{2}+n)}{\Gamma(n+1)\Gamma(\tfrac{3}{2}-n)}\tanh^{n}\tfrac{\eta}{2}\,
F(-\tfrac{1}{2},\tfrac{3}{2};1+n;-\sinh^2\tfrac{\eta}{2})\,.
\end{equation}
Then using that for $\tau_0=i$ we have
$\tanh^{2}\tfrac{\eta}{2}=e^{-2\pi\text{Im}\,\mathcal{T}}$ and
$n=2m$ the result Eq. \eqref{leadingorderf} follows.}

\begin{align}
    f(\mathcal{T},\bar{\mathcal{T}})\approx & \sum_{(p,n)\neq(0,0)}\frac{1}{(p^2+n^2)^{3/2}}\nonumber\\
    &+\sum_{m=1}^{\infty}\sum_{(p,n)\neq(0,0)}\frac{e^{2mi\beta(p,n;i)}}{(p^2+n^2)^{3/2}}
\frac{\Gamma(\tfrac{3}{2}+2m)}{\Gamma(2m+1)\Gamma(\tfrac{3}{2})}
    \,\left(e^{2\pi mi\mathcal{T}}+e^{-2\pi mi\bar{\mathcal{T}}}\right)\,.\label{leadingorderf}
\end{align}

The form of the sum over $m=1,2,\ldots$ in Eq. (\ref{leadingorderf})
which is analogous to the D-instanton case prompts us to assume that
it reproduces the contribution of single multiply--charged Q- and
anti-Q-instantons as one can see by comparing (\ref{leadingorderf})
with Eq. \eqref{saddleapproxwithimagpart}. The first term in
\eqref{leadingorderf} does not correspond to an instanton
contribution. Its origin is yet to be understood.

We have discussed in detail how to obtain the Fourier series
expansion of the function $f$ around $\tau_0=i$, Eq.
\eqref{finalresultfori}. We end this Section by briefly discussing
the Fourier series expansion of $f$ around $\tau_0=\rho$. The
starting point is Eq. \eqref{masterformulaFouriercoefficients} in
which we take $\tau_0=\rho$ and
$\sqrt{\text{det}\,Q}=\tfrac{\pi}{3}$ see table
\ref{tableorbifolddata}. In this case from Eqs. \eqref{cosbeta} and
\eqref{sinbeta} we can obtain the following three identities

\begin{align}
    &\beta(n,n-p;\rho)=\frac{2\pi}{3}+\beta(p,n;\rho)\,,\label{shiftby2thirdpi}\\
    &\beta(p-n,p;\rho)=\frac{4\pi}{3}+\beta(p,n;\rho)\,\label{shiftby4thirdpi}\\
    &-\beta(n,p;\rho)=\frac{4\pi}{3}+\beta(p,n;\rho)\,.\label{betaminusbeta}
\end{align}

\noindent Using \eqref{shiftby2thirdpi} and \eqref{shiftby4thirdpi}
one can show, in a way which is very similar to the derivation of
Eq. \eqref{Fouriercoefficientsfori} for $\tau_0=i$, that the Fourier
coefficients $a_{3m}(\eta)$ are given by

\begin{equation}\label{Fouriercoefficientsforrho}
    a_{3m}(\eta)=\frac{1}{2\pi}\sum_{(p,n)\neq(0,0)}
    \frac{(\text{Im}\,\rho)^{3/2}}{\vert p+n\rho\vert^3}e^{3mi\beta(p,n;\rho)}\,
    \int_{0}^{2\pi}d\theta\frac{\cos(3m\theta)}{\left(\cosh\eta+\sinh\eta\cos\theta\right)^{3/2}}\,.
\end{equation}

\noindent It follows by employing Eq. \eqref{betaminusbeta} that
$a_{3m}(\eta)=a_{-3m}(\eta)$. Hence, using the Fourier decomposition
\eqref{Fourierdecomposition} and Eqs.
\eqref{Fouriercoefficientsforrho}, \eqref{integralrepresentation}
and \eqref{functionalrelationF} we find for $\tau_0=\rho$

\begin{align}
    &f(\eta,\varphi)=\sum_{(p,n)\neq(0,0)}\frac{(\text{Im}\,\rho)^{3/2}}{\vert p+n\rho\vert^{3}}\,
    F(\tfrac{3}{4},-\tfrac{1}{4};1;-\sinh^2\eta)+
    \sum_{m=1}^{\infty}\sum_{(p,n)\neq(0,0)}\frac{(\text{Im}\,\rho)^{3/2}}{\vert p+n\rho\vert^{3}}
    e^{3mi\beta(p,n;\rho)}\,\times\nonumber\\
    &\times(-1)^m\frac{1}{2^{3m}}\frac{\Gamma(\tfrac{3}{2}+3m)}{\Gamma(3m+1)\Gamma(\tfrac{3}{2})}\,\sinh^{3m}\eta\,F(\tfrac{3m}{2}+\tfrac{3}{4},\tfrac{3m}{2}-\tfrac{1}{4};3m+1;-\sinh^2\eta)\left(e^{3mi\varphi}+e^{-3mi\varphi}\right)\,.\label{finalresultforrho}
\end{align}

\noindent At leading order we can write

\begin{equation}
    \sinh^3\eta\approx 8\,e^{-2\pi\text{Im}\,\mathcal{T}}\,,
\end{equation}

\noindent so that at this order near $\tau_0=\rho$ we obtain

\begin{align}
    &f(\mathcal{T},\bar{\mathcal{T}})\approx\sum_{(p,n)\neq(0,0)}\frac{(\text{Im}\,\rho)^{3/2}}{\vert p+n\rho\vert^{3}}+\sum_{m=1}^{\infty}\sum_{(p,n)\neq(0,0)}\frac{(\text{Im}\,\rho)^{3/2}}{\vert p+n\rho\vert^{3}}e^{3mi\beta(p,n;\rho)}\,\times\nonumber\\
    &\times(-1)^m\frac{\Gamma(\tfrac{3}{2}+3m)}{\Gamma(3m+1)\Gamma(\tfrac{3}{2})}\,\left(e^{2\pi mi\mathcal{T}}+e^{-2\pi mi\bar{\mathcal{T}}}\right)\,,\label{leadingorderrho}
\end{align}

\noindent where we used $\varphi=\tfrac{2\pi}{3}\chi'$.

The expressions \eqref{leadingorderf} for $\tau_0=i$ and
\eqref{leadingorderrho} for $\tau_0=\rho$ can be contrasted with the
leading order result for $\tau_0=i\infty$, Eq.
\eqref{asymptFseries}. The results \eqref{leadingorderf} and
\eqref{leadingorderrho} differ from \eqref{asymptFseries} most
notably in the axion-independent parts. We expect that there to be a
Q-brane interpretation for the $\chi'$ independent pieces of
\eqref{leadingorderf} and \eqref{leadingorderrho}, but at this
moment it is not clear what kind of processes would account for
these terms.

\section{Discussion}\label{sec:discussion}

In this article we have constructed new 1/2 BPS instanton solutions
to the Wick rotated Euclidean IIB supergravity theory. We have shown
that they differ from the known D-instantons and that they are the
electric partners of the Q7-branes of
\cite{Bergshoeff:2006jj,Bergshoeff:2007aa}. The path integral
approach to the Q-instantons shows the existence of new vacua and a
new superselection parameter $\chi'_{\infty}$. Further, we have
argued that the Q-instantons contribute to the $\mathcal{R}^4$ terms
near the points $\tau_0=i,\rho$ of the quantum moduli space
${{PSL(2,\mathbb{R})}\over{SO(2)\times PSL(2,\mathbb{Z})}}$. The
expansion of the generalized Eisenstein series around the points
$\tau_0=i,\rho$ contains terms that do not depend on $\chi'$ and for
which a Q-brane interpretation is yet to be found.

We believe that the results of this article together with
\cite{Bergshoeff:2006jj,Bergshoeff:2007aa} support the idea that IIB
supergravity provides a valid field theory approximation of some
underlying quantum theory near each of the orbifold points of the
quantum axion-dilaton moduli space of Figure
\ref{fundamentaldomain}. It is of interest to understand if in
addition to the Q7-branes and the Q-instantons there exist other
Qp-brane solutions associated to the orbifold points $i$ and $\rho$
of the IIB quantum moduli space.

In addition to the instantons and 7-branes it should also be
possible to consider 3-branes near the orbifold points
$\tau_0=i,\rho$ since the 3-brane is an $SL(2,\mathbb{Z})$ singlet
and can be put at any point of the IIB moduli space. Based on the
arguments presented in this paper we expect there to exist a field
theory description of the world-volume theory of a ``Q3-brane'',
\emph{i.e.} a 3-brane near $\tau_0=i,\rho$. It would then be
interesting to study such a Q3-brane in the presence of probe
Q-instantons or probe Q7-branes and to see if one may learn
something about the Yang--Mills theory on the Q3-branes.

We end this discussion Section with the following comment on the
relevance of Q-branes in relation to gauged supergravities. The idea
that the type IIB supergravity theory can be used as a valid
approximation of some underlying quantum theory near the points
$\tau=i,\rho$ is of importance, for example, if one considers gauged
supergravities that result from the IIB theory in which the
$\chi'\rightarrow\chi'+b$ isometry has been gauged. The simplest
example of such a gauged supergravity is the nine-dimensional
$SO(2)$ gauged maximal supergravity that is constructed via a
Scherk--Schwarz reduction of IIB supergravity by gauging the $SO(2)$
subgroup of $SL(2,\mathbb{R})$ \cite{Bergshoeff:2002mb}. This
nine--dimensional theory has domain-wall solutions which correspond
(via uplifting) to Q7-branes of ten-dimensional IIB
supergravity\footnote{The gauging of the isometry associated with
the shift invariance of the RR axion $\chi$ leads to a
nine--dimensional gauged maximal supergravity with $\mathbb{R}$
gauge group whose domain-wall vacuum is associated with the
D7-brane.}. Thus, the study of the structure of gauged
supergravities may provide us with additional information about the
nature of Q-branes and whether they manifest yet unexplored corners
of M-theory.

\section*{Acknowledgements}

This work is supported by the European Commission FP6 program
MRTN-CT-2004-005104 and by the INTAS Project Grant 05-1000008-7928
in which E.B., J.H., A.P.~ are associated to Utrecht university and
D.S.~is associated to the Department of Physics of Padova
University. The work of E.B.~is partially supported by the Spanish
grant BFM2003-01090. J.H. is supported by a Breedte Strategie grant
of the University of Groningen. Work of A.P. is part of the research
programme of the ``Stichting voor Fundamenteel Onderzoek van de
Materie'' (FOM). Work of D.S. was partially supported by the INFN
Special Initiatives TS11 and TV12 and by the MIUR Research Project
PRIN-2005023102. J.H. wishes to thank the IFT of the University
Aut\'{o}noma in Madrid for its hospitality and financial support
during late stages of this research.

\providecommand{\href}[2]{#2}\begingroup\raggedright\endgroup

\end{document}